\documentclass[preprint,12pt]{elsarticle}

\usepackage{hyperref}
\usepackage{amssymb}

\usepackage{listings}
\usepackage[caption=false]{subfig}
\usepackage{multirow}
\usepackage[]{threeparttable}
\usepackage{graphicx,tabularx,multirow,appendix}
\usepackage{booktabs}
\usepackage{float}
\usepackage[normalem]{ulem}
\usepackage[usenames, dvipsnames]{color}
\usepackage{doi}
\usepackage{url}
\usepackage[colorinlistoftodos]{todonotes}
\usepackage{lmodern}

\definecolor{backcolour}{rgb}{0.95,0.95,0.92}
\definecolor{codegreen}{rgb}{0,0.6,0}
\journal{NUCL INSTRUM METH A}

\begin{document}

\begin{frontmatter}



\title{Interfacing Geant4, Garfield++ and Degrad for the Simulation of Gaseous Detectors}


\author[add1,add2]{Dorothea~Pfeiffer\corref{cor1}}
\ead{dorothea.pfeiffer@cern.ch}
\author[add3]{Lennert~De~Keukeleere}
\author[add4]{Carlos~Azevedo}
\author[add5]{Francesca~Belloni}
\author[add6]{Stephen~Biagi}
\author[add7]{Vladimir~Grichine}
\author[add3]{Leendert~Hayen}
\author[add8]{Andrei R. Hanu}
\author[add9]{Ivana H\v{r}ivn\'{a}\v{c}ov\'{a}}
\author[add2,add10]{Vladimir~Ivanchenko}
\author[add11,add12]{Vladyslav~Krylov}
\author[add2]{Heinrich~Schindler}
\author[add2,add13]{Rob~Veenhof}

\address[add1]{European Spallation Source (ESS AB),P.O. Box 176, SE-22100 Lund, Sweden}%
\address[add2]{CERN, CH-1211 Geneva 23, Switzerland}%
\address[add3]{Instituut voor Kern- en Stralingsfysica, KU Leuven, Belgium}%
\address[add4]{I3N - Physics Department, University of Aveiro, 3810-193 Aveiro, Portugal}%
\address[add5]{CEA Saclay, 91191 Gif-sur-Yvette, France}%
\address[add6]{Department of Physics, University of Liverpool, UK}%
\address[add7]{Lebedev Physical Institute of RAS, Moscow, Russia}%
\address[add8]{NASA Goddard Space Flight Center, Greenbelt, Maryland 20771, USA}%
\address[add9]{Institut de Physique Nucl\'{e}aire, Universit\'{e} Paris-Sud, CNRS-IN2P3, Orsay, France}%
\address[add10]{Tomsk State University, Tomsk, Russia}%
\address[add11]{Laboratoire de l'Acc\'{e}l\'{e}rateur Lin\'{e}aire (LAL), Universit\'{e} Paris-Sud XI, CNRS/IN2P3, 91898 Orsay, France}%
\address[add12]{Taras Shevchenko National University of Kyiv (TSNUK), Kyiv, Ukraine}%
\address[add13]{Uludag University, 16059 N{\"u}lufer-Bursa, Turkey}%

\cortext[cor1]{Corresponding author}

\cortext[cor1]{Corresponding author}
\begin{abstract}
For several years, attempts have been made to interface Geant4 and other software packages with the aim of simulating the complete response of a gaseous particle detector. In such a simulation, Geant4 is always responsible for the primary particle generation and the interactions that occur in the non-gaseous detector material. Garfield++ on the other hand always deals with the drift of ions and electrons, amplification via electron avalanches and finally signal generation. For the ionizing interaction of particles with the gas, different options and physics models exist. The present paper focuses on how to use Geant4, Garfield++ (including its Heed and SRIM interfaces) and Degrad to create the ionization electron-ion pairs in the gas. Software-wise, the proposed idea is to use the Geant4 physics parameterization feature, and to implement a Garfield++ or Degrad based detector simulation as an external model. With a Degrad model, detailed simulations of the X-ray interaction in gaseous detectors, including shell absorption by photoelectric effect, subsequent Auger cascade, shake-off and fluorescence emission, become possible. A simple Garfield++ model can be used for photons (Heed), heavy ions (SRIM) and relativistic charged particles or MIPs (Heed). For non-relativistic charged particles, more effort is required, and a combined Geant4/Garfield++ model must be used. This model, the Geant4/Heed PAI model interface, uses the Geant4 PAI model in conjunction with the Heed PAI model. Parameters, such as the lower production cut of the Geant4 PAI model and the lowest electron energy limit of the physics list have to be set correctly. The paper demonstrates how to determine these parameters for certain values of the W parameter and Fano factor of the gas mixture. The simulation results of this Geant4/Heed PAI model interface are then verified against the results obtained with the stand-alone software packages.

\end{abstract}


\end{frontmatter}


\section{Introduction}
\label{introduction}
Geant4 (GEometry ANd Tracking)~\cite{Geant4a, Geant4b, Geant4c} is an object-oriented C++ toolkit for the simulation of the passage of particles through matter. In particle physics, Geant4 is the most commonly used software package for Monte Carlo simulations. However, its application areas also include nuclear and accelerator physics, as well as studies in medical and space science.  In recent years, the framework has been extended to include low energy applications, and several extensions from different fields have been added. The NXSG4 package~\cite{Geant4_Diffraction}, for example, adds functionality for Polycrystalline Neutron Scattering in Geant4, whereas the DNA project~\cite{DNA1,DNA2} extends Geant4 with processes for the modeling of early biological damage induced by ionizing radiation at the DNA scale.

This paper will focus on the simulation of gaseous detectors and presents an approach to interface Geant4 with Garfield++ and Degrad, using the Geant4 parameterization feature~\cite{Parametrization}. The Geant4/Garfield++ interface, as described in this paper, has been integrated into the European Spallation Source Geant4 framework~\cite{Geant4_ESS}, and into Geant4 VMC~\cite{VMC}. In Geant4, the G4PAIModel (photo absorption ionization)~\cite{PAI} and the G4PAIPhotonModel (photo absorption ionization photon model)~\cite{paiph} were designed for the simulation of gaseous detectors. Both models are derived from the work of Allison and Cobb~\cite{Cobb} and emerge from the assumption that soft collisions, in which a virtual photon is absorbed by the atom as a whole, dominate the interactions of relativistic particles with a gas. As a result, the calculated energy loss in these models is mainly determined by the photon absorption cross-section of the individual atoms. In addition, it gives rise to two main regions with respect to the energy of the ionization electrons: the resonance region, where low-energy 'conduction' electrons are created, and the pseudo-free region, in which the ionization electron receives a large amount of energy and the kinematics can be approximated through Rutherford scattering on a nearly free electron~\cite{Cobb}. The difference between the G4PAIModel and the G4PAIPhotonModel is that the latter also produces photons. Although the resulting differences are not substantial, the G4PAIPhotonModel more accurately describes the spatial charge distribution~\cite{paiph}\footnote{Unless explicitly stated, the term PAI model in this paper stands for both G4PAIModel and G4PAIPhotonModel models.}.

Traditionally, the Geant4 PAI model was designed for the transport of fast charged particles in thin absorbers~\cite{PAI_thinAbsorbers}, but has recently been extended to include low energy primary particles~\cite{PAI_extension}. Since Geant4 version 10.2, atomic deexcitation (fluorescence, Auger electron emission including Auger cascades and particle-induced X-ray emission) can be activated for electromagnetic physics processes like the photoelectric effect, ionization and Compton scattering. A simulation using the PAI model is thus able to describe the number and positions of the initial ionization electron-ion pairs, but it does still not include additional processes like attachment, recombination, drift and diffusion. Furthermore, although electromagnetic fields are implemented in Geant4, it is not possible to simulate the process of an electron avalanche or the associated signals induced on pick-up electrodes or wires at high voltage. This emphasizes the need for an interface between Geant4 and a software package that can simulate the above-mentioned features of gaseous detectors, such as wire chambers, Resistive Plate Chambers (RPCs) or Micro Pattern  Gaseous Detectors (MPGDs).

Dedicated to the simulation of gaseous detectors such as Micro Pattern Gaseous Detectors (MPGD) ~\cite{MPGD} is the Fortran software package Garfield~\cite{Garfield}. The software has been ported to C++ by the name of Garfield++~\cite{Garfieldpp}, and makes extensive use of the ROOT framework~\cite{Root} for data analysis and visualization. For several years, attempts have been made to interface Geant4 and Garfield~\cite{FortranInterface}, a task which has been now hugely facilitated by the availability of Garfield++. Garfield++ accepts two- and three-dimensional field maps computed by various finite element programs as a basis for its calculations of drift and avalanche processes. The finite element technique can handle complex electrode shapes as well as dielectrics. For the computation of electron transport properties in nearly arbitrary gas mixtures, Garfield++ includes an interface to Magboltz~\cite{Magboltz}. Furthermore, Garfield++ also includes interfaces to Heed, a PAI model implementation~\cite{Heed_PAI} similar to that in Geant4, and SRIM (Stopping and Range of Ions in Matter)~\cite{SRIM}. Both can be used to create the initial ionization electron-ion pairs. 

Even more details concerning ionization, excitation levels and vibrational modes can be obtained by Degrad, another Fortran program~\cite{degradWeb}. Degrad includes an accurate Auger cascade model for the interaction of photons, electrons and ionizing particles with gas mixtures in electric and magnetic fields. For X-rays, the software automatically simulates shell absorption by the photoelectric effect, Compton scattering or pair-production and the subsequent Auger, Coster-Kronig, shake-off and fluorescence emission. Bremsstrahlung emissions by secondary electrons are also included \cite{degradRef}. With regard to the included physics processes, Degrad is hence the most complete software package to simulate the interaction of photons or electrons with various gases.

In the following, the paper first describes the different scenarios to interface Geant4 with Garfield++ (and its interfaces to Heed and SRIM) and Degrad in section~\ref{Division of tasks}. In section~\ref{Implementation}, the basic software steps for the interface are explained using two simulation examples: a Xenon-based TPC (section~\ref{Simulation example: Xenon-based optical TPC}) and a TPC with charge sensitive readout (section ~\ref{Simulation example: TPC with charge sensitive readout}). Section~\ref{Geant4 simulation parameters and their optimization} discusses how to optimize the key simulation parameters in Geant4, the lower production cut and the lowest electron energy limit, with the help of the W value and the Fano factor. All simulation scenarios besides the Geant4/Heed PAI model interface (option D in section~\ref{Division of tasks}) utilize just one physics model. But since the Geant4/Heed PAI model interface combines the Geant4 PAI model and the Heed PAI model, a verification of the simulation results is needed. Section~\ref{Verification} illustrates how the simulation results depend on the tuning of the Geant4 lower production cut and the transfer energy threshold. The obtained simulations results (deposited energy spectra, spatial distribution of electron-ion pairs, simulation time) are then compared with the results of the stand-alone programs. Finally, in section ~\ref{Conclusion}, the paper results are summarized.

\section{Division of tasks between Geant4, Degrad and Garfield++}
\label{Division of tasks}
\begin{figure}[htbp]
 \centering
 \includegraphics[height=10cm]{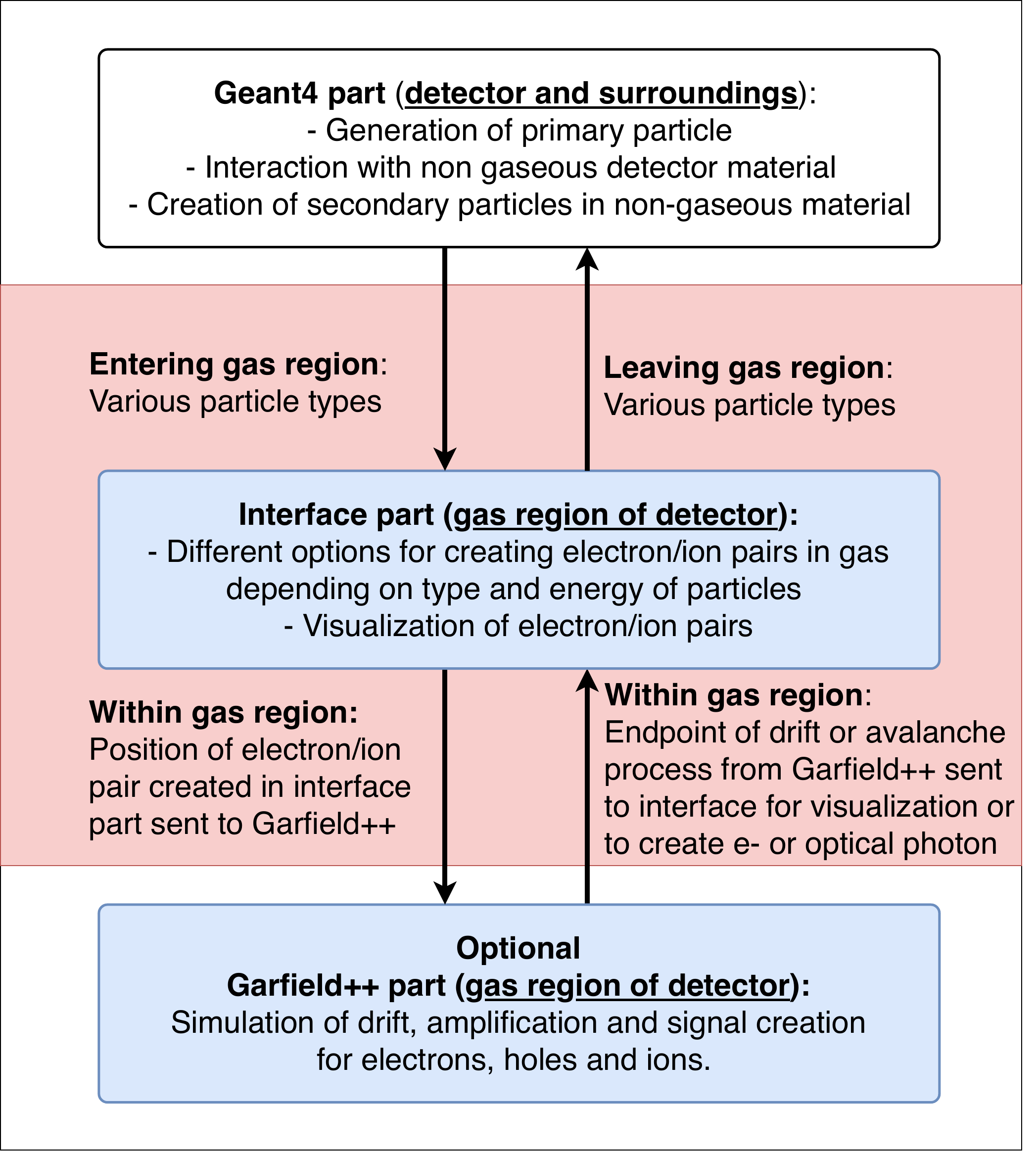}%
 \caption{Division of simulation tasks between Geant4, the interface part (marked with a red background) and Garfield++. The main topic of this paper is the interface part, for which different options exist.}
 \label{fig: task_division}
\end{figure}

In the present section, possible scenarios for interfacing the different software packages will be summarized and discussed. Next, implementation details will be described with the help of a detailed example. Figure~\ref{fig: task_division} shows a schematic overview of the general simulation setup. Features like the interaction of high-energy particles with matter, or complex detector configurations containing various non-gaseous materials, can neither be implemented in Garfield++ nor Degrad. Geant4, shown in the white box at the top of the flow chart, is thus always used for the primary particle generation, the interaction of the primary particle with the detector material, and the possible creation of secondary particles in the detector material. The gaseous regions of the detector are displayed in blue. Garfield++ on the bottom always deals with the drift of ions and electrons, amplification via electron avalanches and finally signal generation. 

As part of the complete simulation, the paper will concentrate in the following on the interface part in the gas region. The interface part, marked with a red background, is describing the ionizing interaction of particles with the gas, for which different physics models and software packages exist. The simulation result of the interface part always consists of positions and production times of electron-ion pairs~\footnote{Throughout the paper, the term electron-ion pairs refers to all electron-ion pairs produced when transport is done down to energies below the lowest ionization potential in the gas phase, or to thermal electron energies, depending on the case.}. In the Garfield++ part of the simulation, the user can then simulate drift or avalanche processes with these electron or ion positions. The endpoints of the drift or avalanche processes in Garfield++ can be transferred back to the interface part, \textit{e.g.} to create electrons or optical photons in Geant4. Further, the positions of the electrons during the drift or avalanche process can be sent to the interface part. The standard Geant4 visualization can then be used to display drift lines or avalanches. Due to the whole simulation being contained in a single C++ program, all the involved software components can easily exchange data. Nevertheless, Geant4, Garfield++ and Degrad remain separate programs. A Geant4 particle track, \textit{e.g.}, cannot be directly handled by Garfield++, since the programs use different particle classes. The exchange between Geant4, Garfield++ and Degrad works thus in both directions based on the particle properties like positions and kinetic energies.

\begin{figure}[htbp]
\centering
\subfloat[All particle types\label{fig: all_particles}]{
\includegraphics[height=6cm]{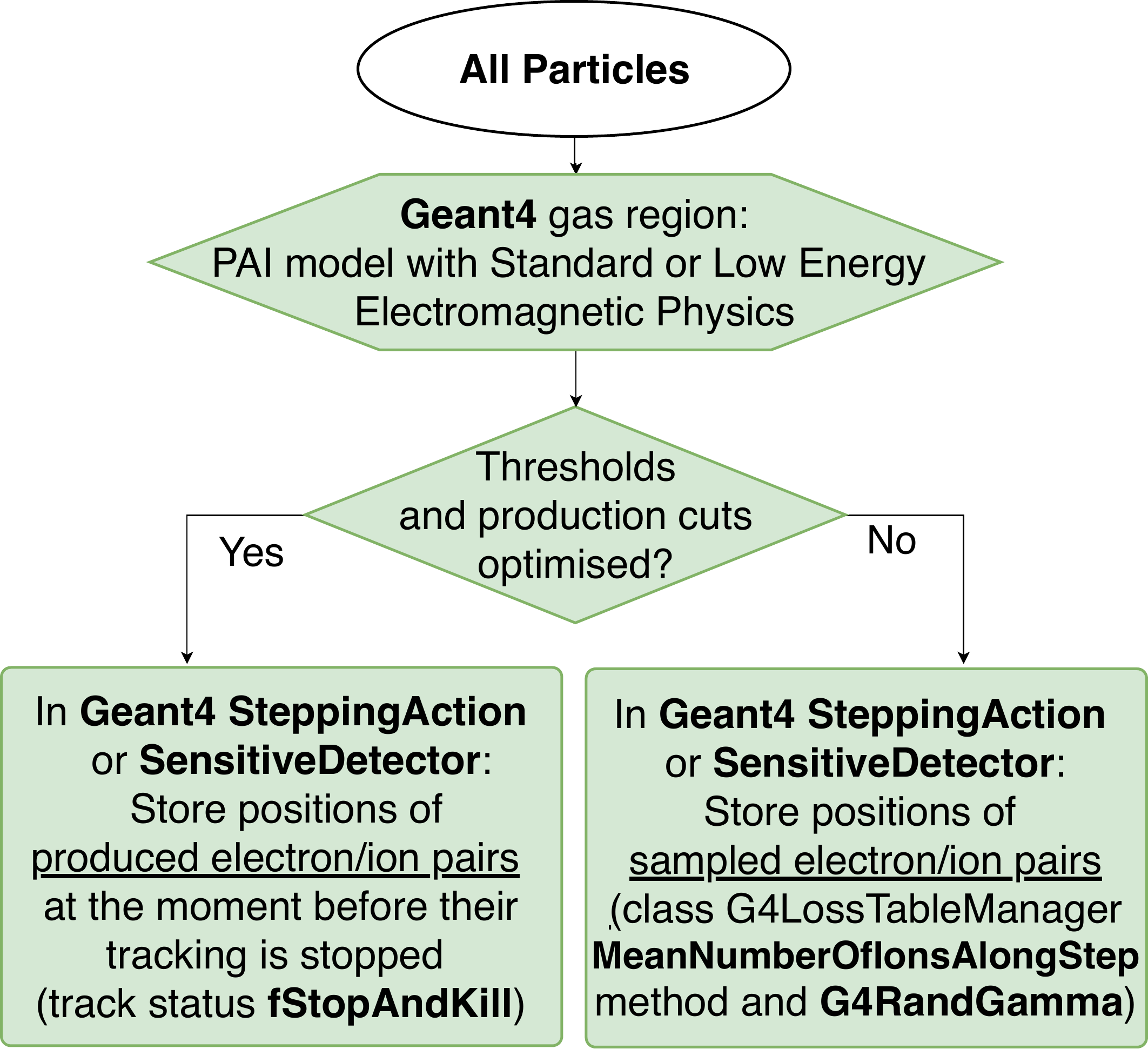}%
}
\subfloat[Electrons\label{fig: electrons}]{
\includegraphics[height=6cm]{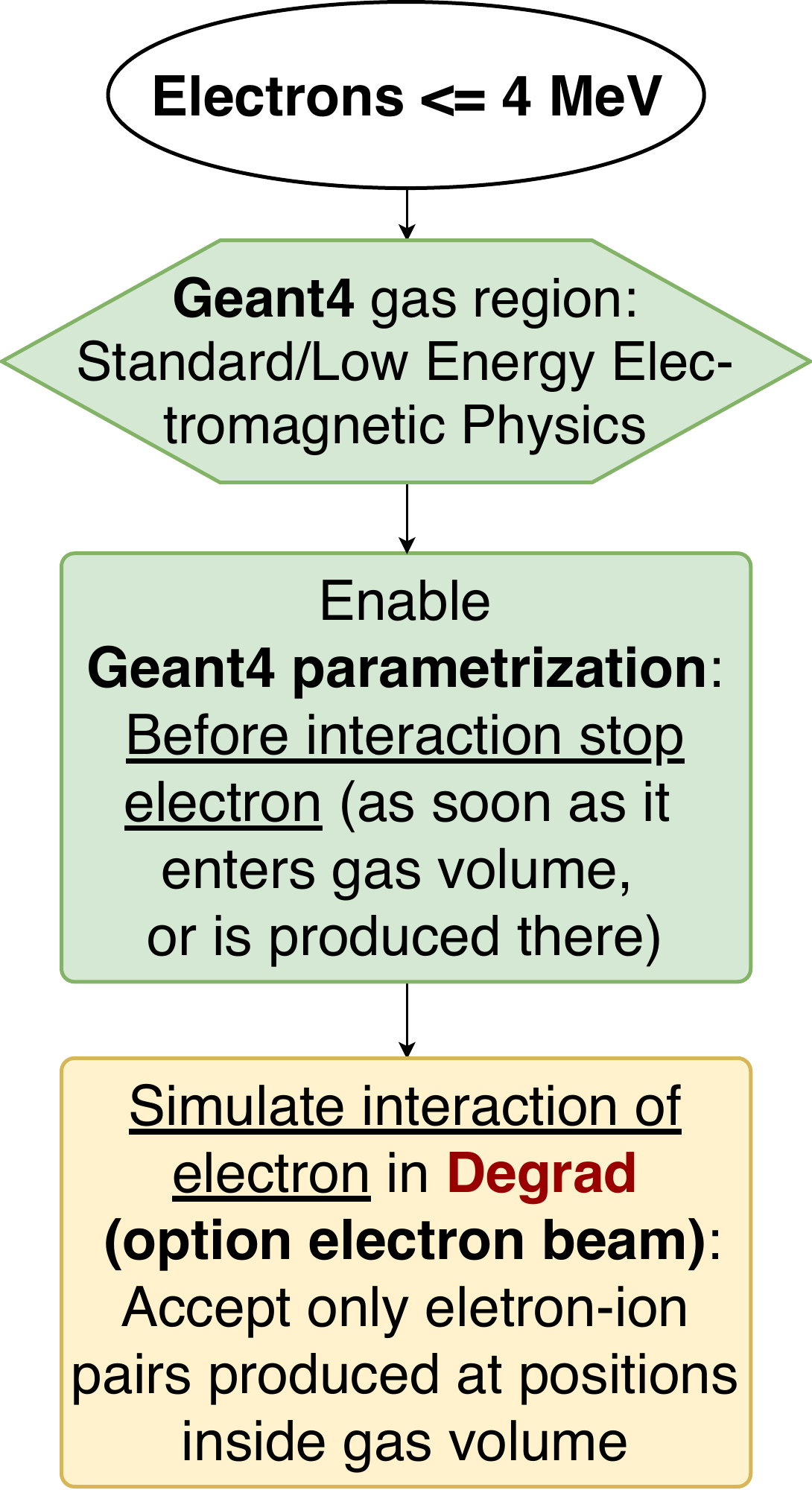}%
}
\subfloat[Ions\label{fig: ions}]{
\includegraphics[height=6cm]{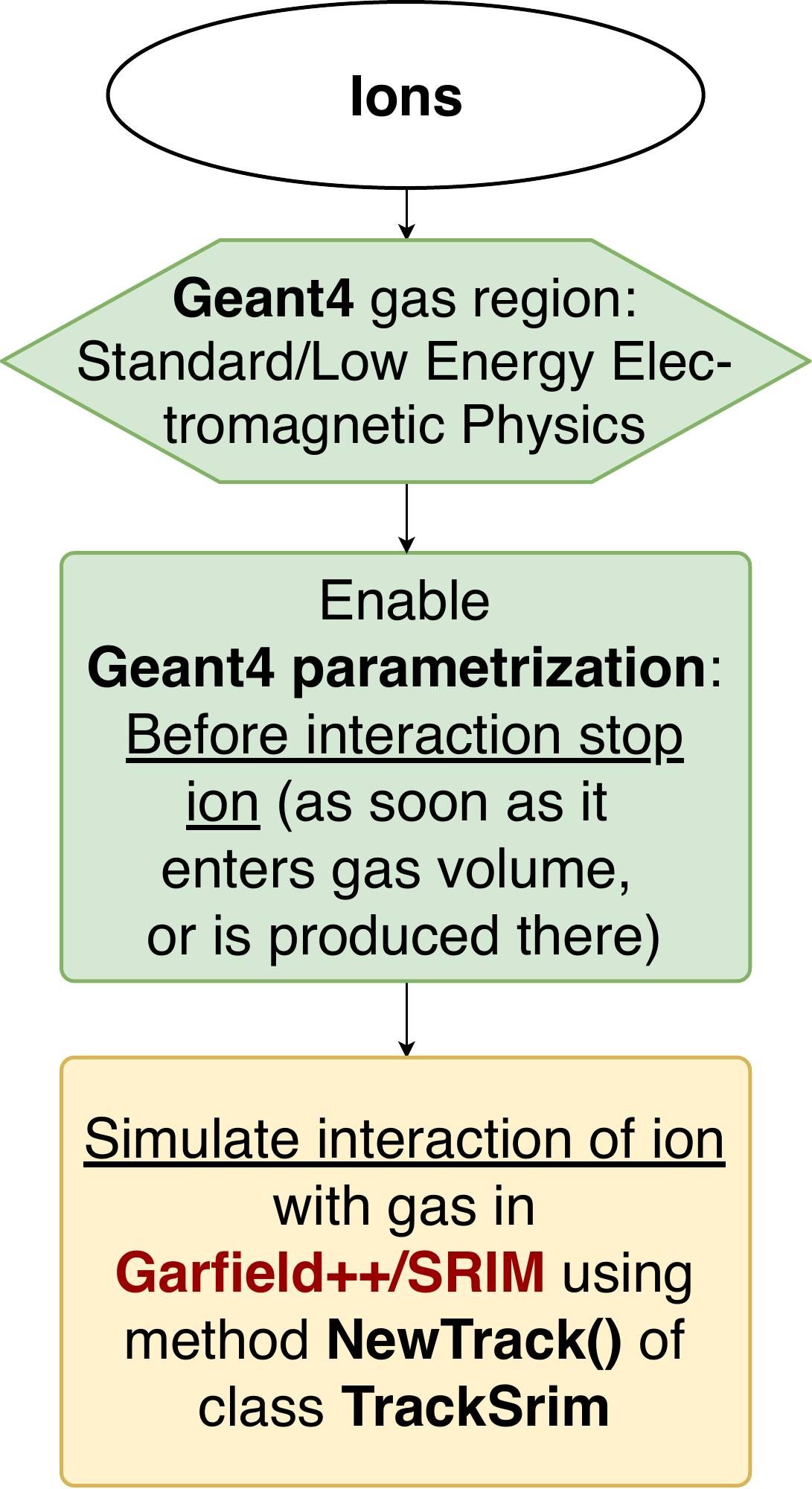}%
}\\
\centering

\subfloat[Photons\label{fig: photons}]{
\includegraphics[height=7cm]{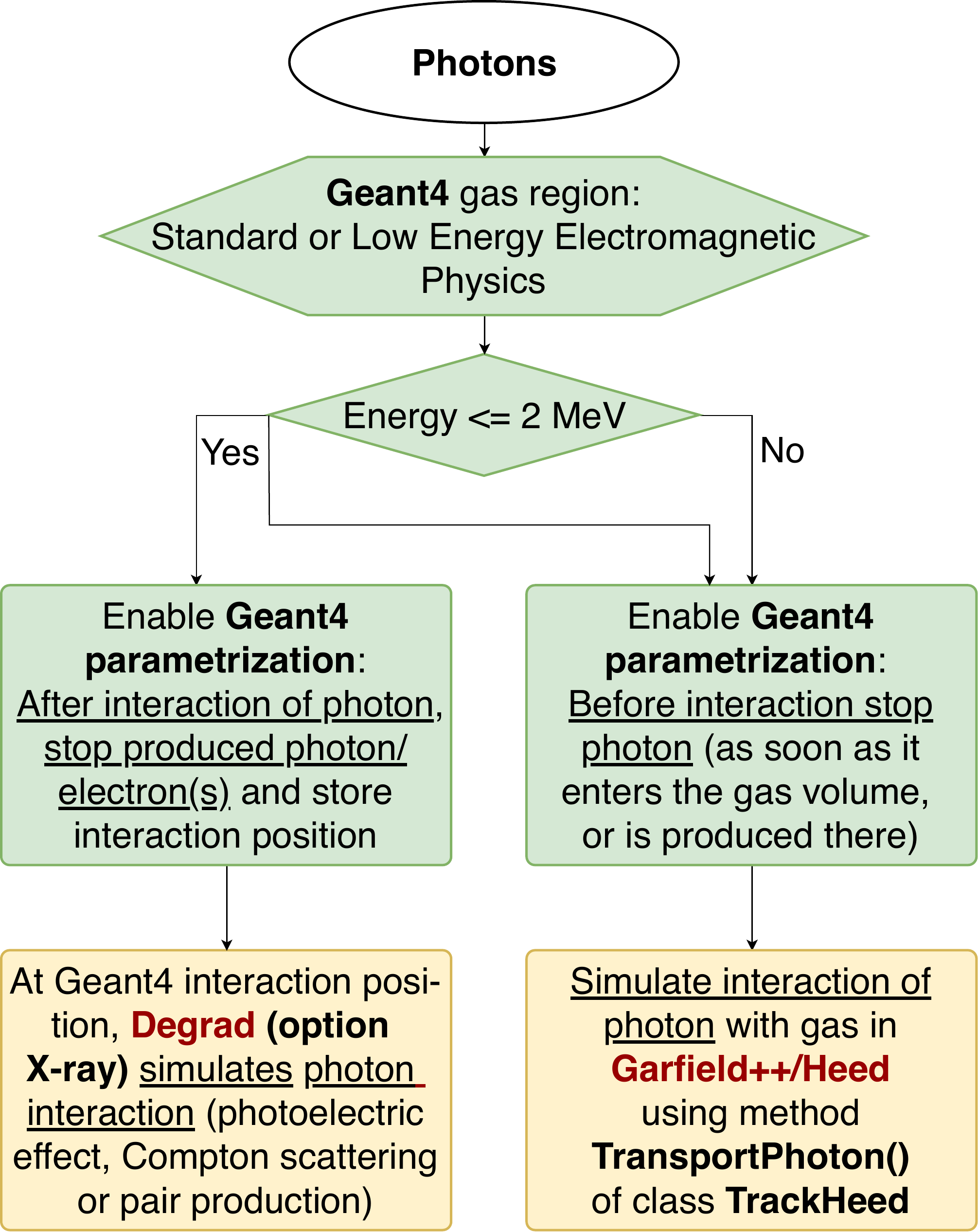}%
}
\subfloat[Charged particles\label{fig: charged_particles}]{
\includegraphics[height=7cm]{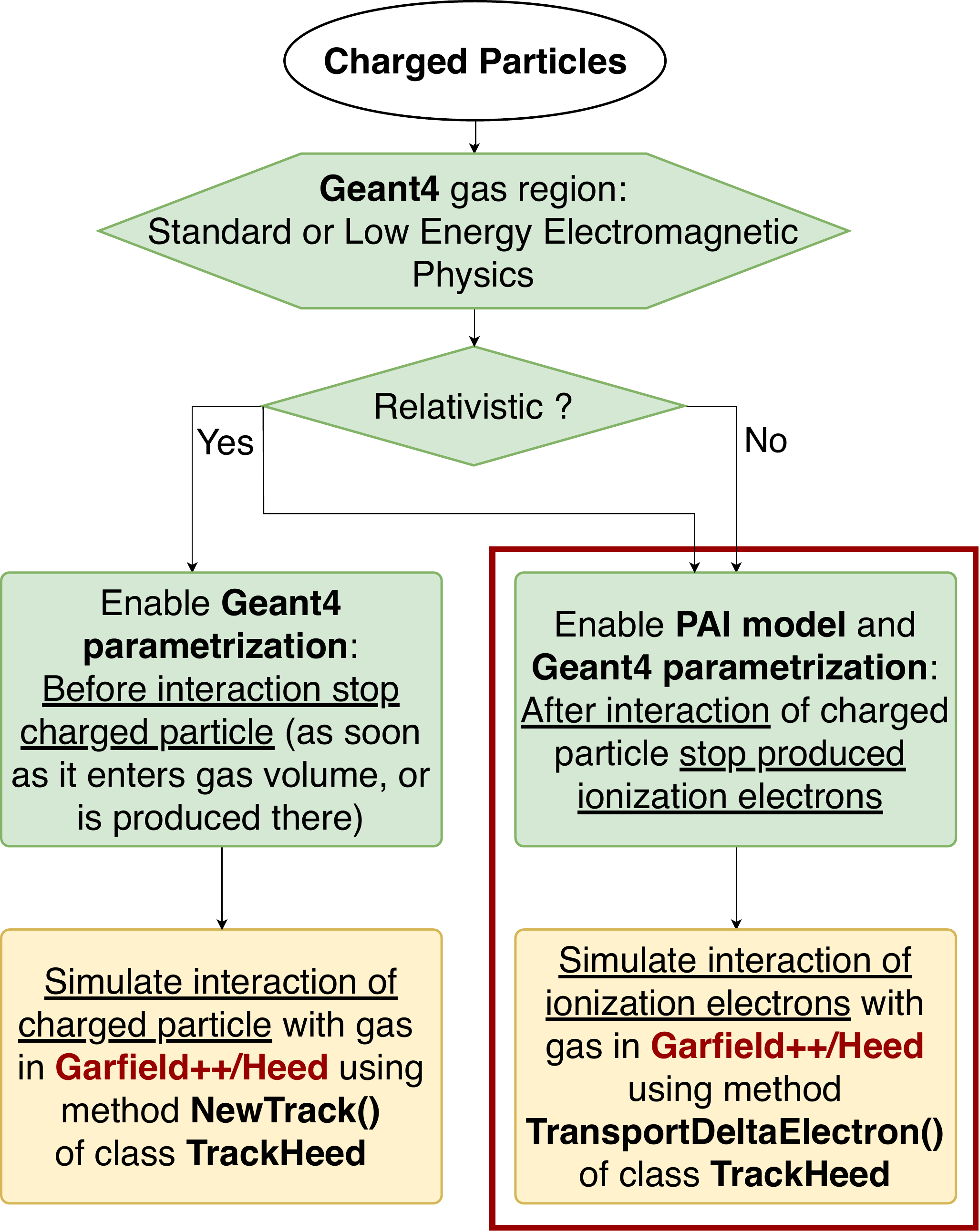}%
}\\
\caption{The diagrams illustrate the different interface options depending on particle type and kinetic energy. The green elements represent steps taking place in Geant4, yellow elements indicate actions taking place in Garfield++ (Heed or SRIM) and Degrad. The method the user has to call is printed in bold letters. The red box in the flow chart path for charged particles indicates that this is the only case, where two different physics models, Geant4 and Heed PAI model, are used in conjunction for the generation of electron-ion pairs.}
\label{fig:all_particles}
\end{figure}

Through the interface part, there are hence different options to create the electron-ion pairs in the gas phase, besides the native Geant4 ones (discussed first):

\begin{itemize}
\item{A) Geant4 production of electron-ion pairs (all particle types)\\
This option is shown in the left part of the flow chart in figure~\ref{fig: all_particles}. Careful tuning of the lowest electron energy limit and the lower production cut is needed to produce enough electron-ion pairs. The interplay between these two parameters is difficult to predict, as explained in detail in section~\ref{Lower production cut and lowest electron energy limit}. The tuning might not be possible in all cases, but works for charged particles in combination with the PAI model. To obtain the correct number and positions of the electron-ion pairs, it is essential to capture these positions just before the tracking of the particle in the gas is stopped. This way one can be sure that no further ionization occurs. The final track position is thus stored in the user SteppingAction or SensitiveDetector methods. In each Geant4 simulation, one of these two methods has to be implemented to access particle information at each simulation step. Due to the need for obtaining the final position, the Geant4 physics parametrization feature, which works only with particles that are still being tracked, cannot be used easily. The user has to write the necessary source code that takes care of the transfer of the electron-ion pair positions to Garfield++.}

\item{B) Geant4 sampling of electron-ion pairs (all particle types)\\
As an alternative, if high precision is not required, the numbers and positions of the electron-ion pairs can also be sampled from the energy deposition in each step in a SteppingAction or SensitiveDetector, without actually creating new electrons. This option is shown in the right part of the flow chart in figure~\ref{fig: all_particles}. With the sampling, only the value W (mean energy needed to create an electron-ion pair via ionization) can be correctly reproduced, but not the Fano factor, as it will be shown in section~\ref{Verification}. As in case A, the Geant4 physics parametrization feature cannot be used easily, since the parametrization requires actual electrons, and does not work for sampled positions.  }

\item{C) Heed  (charged particles with relativistic energies)\\
To simulate the interaction of relativistic particles (protons, muons, electrons, pions, etc), the PAI model of Heed can be chosen, as displayed on the left side of figure~\ref{fig: charged_particles}. The Geant4 parametrization hands the control over to Heed as soon as the charged particle enters the gas volume or is produced there. The Heed PAI model implementation uses two methods. To transport fast charged particles, NewTrack() has to be called. NewTrack() tracks the particles without Coulomb scattering, in contrast to the PAI model in Geant4. In a second step, Heed calls then automatically TransportDeltaElectron() for the $\delta$ electrons that were produced by NewTrack(). In contrast to NewTrack(), TransportDeltaElectron() includes Coulomb scattering. Since Heed uses NewTrack() first, the stopping power of the primary is continuous and based on its initial energy, which works fine for Heed's intended purpose of tracking relativistic charged particles in thin absorbers. However, for slower particles in thicker absorbers, it leads to unrealistically straight tracks and an incorrect energy loss\footnote{The qualifiers thin and thick have to be interpreted here with respect to the energy loss of the particle. As soon as the particle loses a substantial amount of energy in the absorber, it cannot be considered as thin anymore.}. Since Heed is not multithread-processor safe, only the Geant4 sequential mode can be used in combination with Heed. }

\item{D) Geant4/Heed PAI model interface (charged particles of all kinetic energies)\\
The right path of the flow chart in figure~\ref{fig: charged_particles} applies to all charged particles, independently from their kinetic energy. The Geant4 parametrization feature and the Geant4 PAI model are activated in the gas region. The Geant PAI model is used to create the primary electron-ion pairs in the gas, at which point Heed takes over. Heed is used to propagate all ionization electrons with a kinetic energy of a few keV and lower as $\delta$-electrons. These $\delta$-electrons have a certain probability to further ionize the gas, until all electrons reach thermal energies~\footnote{Instead of Heed, Degrad could also be used for the $\delta$-electron transport. But due to the faster simulation time and easier use of Heed (it is more straightforward to interface a C++ program than a Fortran program like Degrad), and the similarity of the results obtained in section~\ref{W value and Fano factor}, the authors prefer Heed. If Degrad is used, the option electron-beam (total absorption) has to be chosen. The procedure is equivalent to the description under case E.}. The idea of option D is thus to use the Geant4 PAI model to replace the NewTrack() method of the Heed PAI model, and just use the TransportDeltaElectron() method of Heed.  In this way, the complete tracking uses Coulomb scattering. But since the Geant4 and the Heed PAI model are used together here, a verification of the results is necessary. The verification can be found in section~\ref{Verification}}. 

\item{E) Degrad (electrons up to 4 MeV)\\
Figure~\ref{fig: electrons} illustrates the use of the Degrad option electron-beam (total absorption). Degrad can be used to simulate the interaction of electrons with kinetic energies of up to 4 MeV. As soon as the electron enters the gas volume or is produced there, the control is handed over from Geant4 to Degrad. Degrad then simulates the interaction of the electron with the gas. The electron is tracked until it reaches the kinetic energy the user set as thermalization energy. Degrad uses an infinite gas volume for the simulation. In the case of ~\textit{e.g.} an MPGD with a small gas volume, many produced electron-ion pairs will have to be discarded, since their positions are outside of the detector volume defined in Geant4.}

\item{F) Degrad (photons up to 2 MeV)\\
The use of Degrad is shown on the left of figure~\ref{fig: photons}. In contrast to Heed, Degrad simulates the shell absorption by photoelectric effect and subsequent Auger, shake-off and fluorescence emission. The program calculates the number of electrons and excitations after energy thermalization and gives the Fano factor for both the electrons and the excitations. Degrad does neither determine the cross-section of the interaction, nor the correct location of the interaction. Degrad assumes that the photon always interacts, and that the interaction takes place at the center of its coordinate system. Therefore, Geant4 is used to determine the cross-section and the location of the interaction. In case the photon interacts, Geant4 creates, depending on the interaction (photoelectric effect, Compton scattering or pair production), either a photoelectron, a Compton photon and electron or two electrons. The position of creation is stored. Subsequently, the model stops the tracking of the produced particles in Geant4 (the particles are \textit{killed}), and hands the control over to Degrad. Degrad now calculates the process interaction probability,~\textit{i.e.} between photoelectric effect, Compton scattering or pair production, and simulates this interaction and subsequent processes like Auger effect and recombination. The disadvantage of this approach is the increased run time of the simulation due to the use of Degrad. Since Degrad always uses a gas volume of infinite size, the user has to check whether the electron-ion pairs are produced within the gas volume defined in Geant4.}

\item{G) Heed (photons)\\
The right part of the flow chart in figure~\ref{fig: photons} shows the use of the Geant4 parametrization feature in combination with Heed for photons. The tracking of the photon is stopped in Geant4 as soon as the photon enters the gas volume, or is produced there. Heed then determines the appropriate cross section for the interaction, and takes care of the creation of electron-ion pairs. The approach works for all photon energies, but the physics implemented in Heed for the transport of the resulting products is limited.}

\item{H) SRIM (ions)\\
Figure~\ref{fig: ions} illustrates the use of the Garfield++ SRIM~\cite{SRIM} interface for ions. With the Geant4 parametrization enabled, the tracking of the ion is stopped in Geant4 as soon as an ion enters the gas volume or is produced there. At the entrance or production position, the interaction of the ion with matter is then simulated with SRIM in Garfield++. The disadvantage of this approach is that the SRIM program has to be run first for each gas composition, particle type and kinetic energy range. SRIM generates a data file containing for different particle energies, the dE/dx, the projected range and longitudinal/lateral straggling. The SRIM interface of Garfield++ then uses this file for its calculations.}

\end{itemize}

\section{Implementation of the interface part}
\label{Implementation}
The code examples that demonstrate the different possibilities described in the paper can be downloaded from Github under:\\
\url{https://github.com/lennertdekeukeleere/Geant4GarfieldDegradInterface}. The repository also includes a step-by-step guide that describes the software implementation of the interface.

\subsection{Software implementation}
\label{Software implementation}
The physics para\-metrization feature in Geant4 offers the easiest way to interface Geant4 with an external software package. The general proposed idea is to create a region, in which a user implementation of the physics and the detector response is provided, instead of the default Geant4 code in this region. This region, defined by the G4Region class, is created during the detector construction, and consists of one or more G4LogicalVolumes, often corresponding to sub-detector volumes. The complete syntax is shown in the class \textbf{DetectorConstruction}. To implement the parameterized physics model, the user has to create a new UserG4Fast\-Simulation\-Model derived from G4VFast\-Simulation\-Model, and attach it to the region. It is possible to attach more than one UserG4Fast\-Simulation\-Model to the same region. The user physics code is now used instead of Geant4 in this region, whereas for the remainder of the geometry, the Geant4 physics is still valid.

In the physics list of the program (class \textbf{PhysicsList}), the fast simulation has to be activated in the AddParameterisation() method. As of Geant4.10.4, the activation is also possible via a macro command. The core part of the interface is the User\-G4Fast\-Simulation\-Model derived from G4VFast\-Simulation\-Model. The name G4VFast\-Simulation\-Model implies that the parameterized model is normally simpler and thus faster than the full Geant4 tracking. In the case of Garfield++ or Degrad, however, the parameterized model is more detailed. The G4VFast\-Simulation\-Model has three pure virtual methods, which must be overridden in the User\-G4Fast\-Simulation\-Model, as \textit{e.g.} shown in class \textbf{HeedModel}. 

The first method, IsApplicable(), must return true when the parametrization model should be applied to the particle under consideration. If this is not the case, the default Geant4 physics will be applied. The second method, ModelTrigger(), is called in every step along the track and should return true if the user-defined conditions of the track are fulfilled. Finally, the implementation of the parameterized model occurs in the DoIt()-method, following one of the scenarios described above in section~\ref{Division of tasks}. More on the implementation of the G4Fast\-Simulation\-Model can be found in the Geant4 User guide for application developers~\cite{Geant4UserGuide}.

In the following, two examples are described to show the power of the interface.

\subsection{Simulation example: Xenon-based optical TPC}
\label{Simulation example: Xenon-based optical TPC}
A $5.9$~keV Fe X-ray is simulated in pure Xenon, using a Degrad model that is valid for electrons with an energy larger than $7$~eV, as shown in class \textbf{DegradModel}. Since Degrad does neither determine the probability/cross-section of the interaction, nor the correct location of the interaction, first Geant4 is used. In case the photon interacts, Geant4 creates, depending on the interaction (photoelectric effect or Compton scattering), either a photoelectron or a Compton photon and electron. The position of creation is stored. Subsequently, the model stops the tracking of the produced particles in Geant4 (the particles are \textit{killed}), and hands the control over to Degrad. Degrad now calculates the process interaction probability,~\textit{i.e.} between photoelectric effect and Compton scattering, and simulates this interaction and subsequent processes like Auger effect and recombination. As Degrad is a Fortran software and writes its output to a result file, the simplest way to transfer information back to Geant4 is based on a file I/O chain: Geant4 runs Degrad with a parameter file and Degrad writes the simulation results to a text file. From this result file, Geant4 reads back the positions and production times of the electron-ion pairs. Degrad always assumes $(x_0,y_0,z_0)=(0,0,0)$ as the interaction position of the photon, and simulates an infinite volume. Therefore, the positions of the electron-ion pairs have to be corrected, using the stored interaction position of the X-ray from Geant4. If the corrected positions are within the gas volume of the detector, the CreateSecondaryTrack() method of the G4FastTrack class is called to create new $7$~eV (the thermalization energy set in Degrad) secondary electrons in Geant4. 

For a simulation of the light production, now a Garfield++ model that describes the simulation of secondary scintillation (electroluminescence) is used (class \textbf{GarfieldVUVPhotonModel}). The Garfield++ model is triggered by the electrons created in the Degrad model, \textit{i.e.} electrons with an energy of $7$~eV. In the model, the AvalancheMicroscopic class in Garfield++ is used to \textit{microscopically} track the electron-ion pairs created after the primary particle interaction, storing a list of the Xe excited states. The SetUserHandleInelastic() method accepts a user-written call-back function, that has access to the time, position and excitation levels in each simulation step. This information is stored in a data structure. In the case of pure noble gases, if one assumes that each excitation will produce a secondary scintillation photon~\cite{Oliveira}, the production of light can be computed in a straightforward way. 

\begin{figure}[htbp]
 \centering
 \includegraphics[width=.7\textwidth]{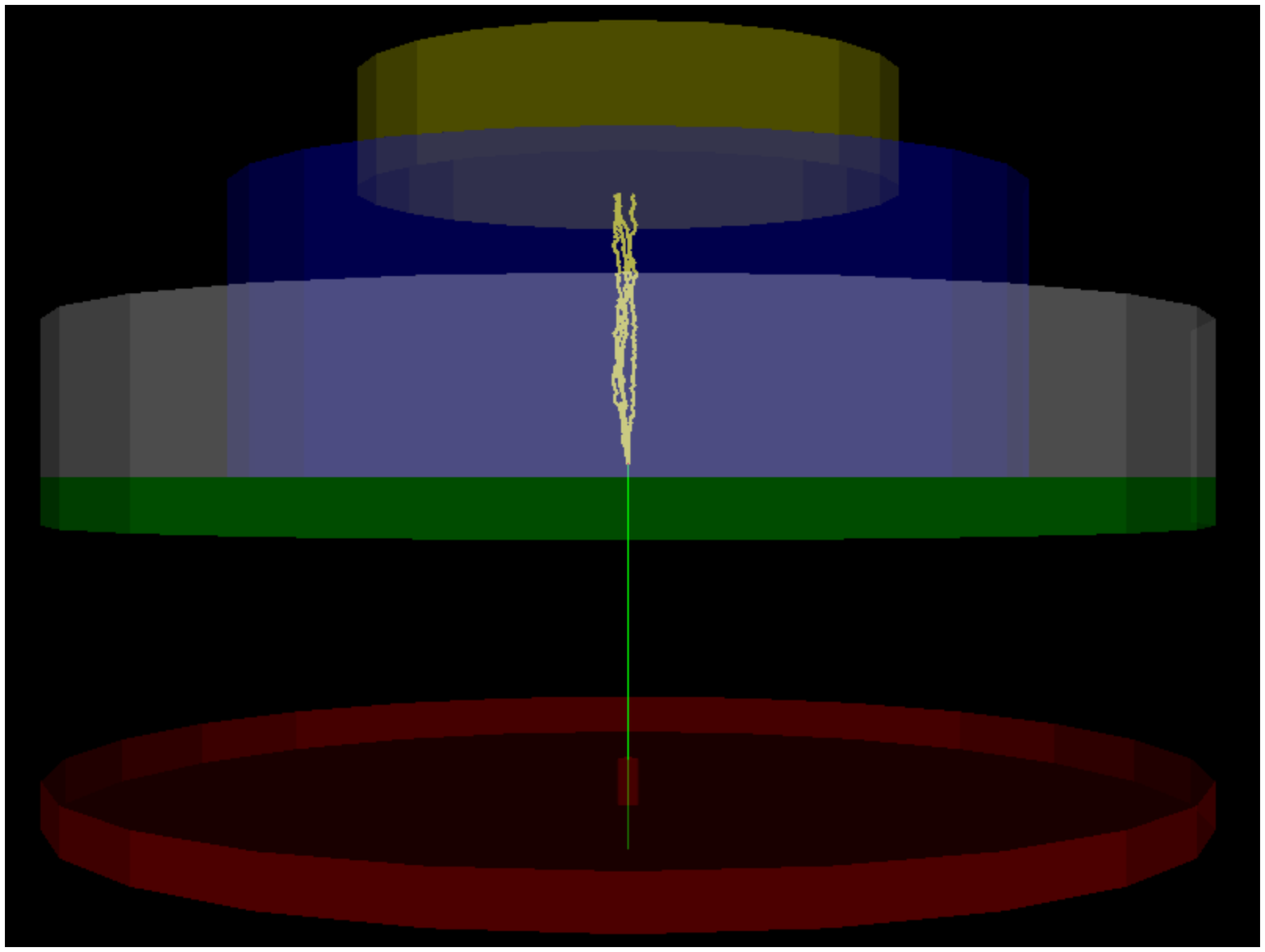}%
 \caption{Xenon-based optical TPC: Electron-ion pairs created by 5.9 keV X-ray (green line) in the Xenon optical TPC. The blue volume is the gas volume, and the yellow cylinder is the PMT used in the example. The electron-ion pairs have been created in Degrad, and are subsequently sent to Garfield++ to drift them along the electrical field lines to the anode (yellow lines). The photons, that are created in the GarfieldVUVPhotonModel are not shown here.}
 \label{fig: Xenon}
\end{figure}

For each element of the data structure filled in the call-back function, a secondary optical photon is produced in Geant4. The optical photon tracking will then be carried out by Geant4. For gas mixtures, a more detailed model should be adopted in Garfield++ in order to include the quenching of the Xenon excimers by the admixture~\cite{Azevedo2017}. Figure~\ref{fig: Xenon} shows the Geant4 visualization of the Xenon-based optical TPC. A 5.9 keV X-ray (green line) created electron-ion pairs in the blue gas volume. These electron-ion pairs were produced in Degrad, and subsequently sent to Garfield++ to drift them along the electrical field lines of the anode. The drift lines are indicated in yellow. Not shown here are the photons, which are created in the \textbf{GarfieldVUVPhotonModel} and will be detected in the yellow PMT.

\subsection{Simulation example: TPC with charge sensitive readout}
\label{Simulation example: TPC with charge sensitive readout}
\begin{figure}[htbp]
\centering
\subfloat[$1$~MeV electron track\label{fig: track}]{
\includegraphics[width=.495\textwidth]{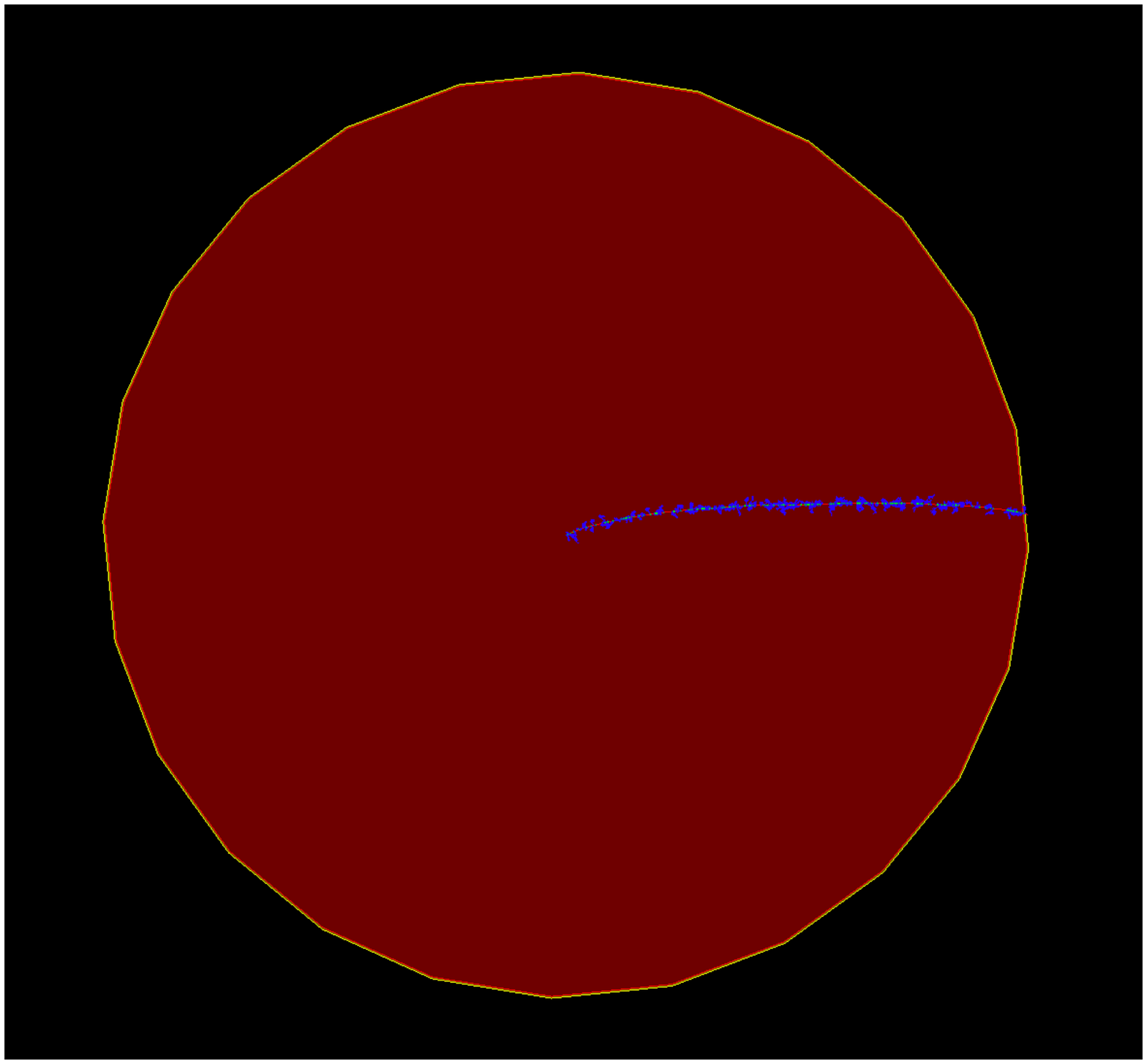}%
}
\subfloat[Ionization electron drift lines\label{fig: drift}]{
\includegraphics[width=.495\textwidth]{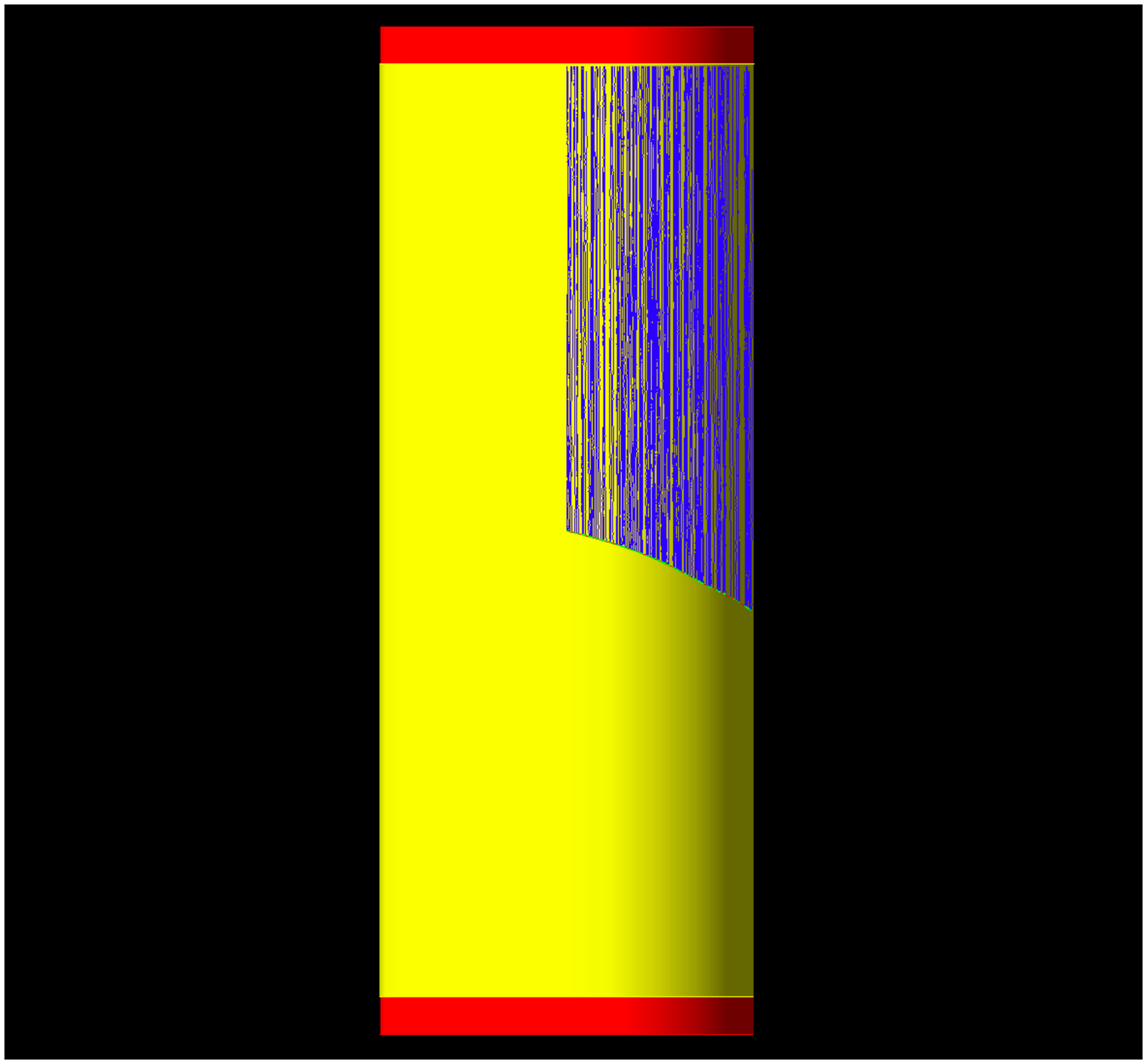}%
}
\caption{Track of a $1$~MeV electron and ionization electron drift lines in a  Ne/CO$_{2}$/N$_{2}$ (90/10/5) TPC. }
\label{fig: Detector}
\end{figure}

This example is based on the Garfield++ ALICE TPC example~\cite{TPC_example}. To be able to also simulate slow electrons, the Geant4/Heed PAI model interface (option D) is used. In this case, the Garfield++ model (class \textbf{HeedInterfaceModel}) is valid for electrons below $1$~keV. Electrons enter the detector, and produce secondary electrons inside the gas volume. The primary particle and the secondary electrons are tracked by the Geant4 PAI model, until their kinetic energy falls under $1$~keV. From this point on, the control is handed over to Heed. Heed treats these electrons then as $\delta$ electrons, and further tracks them using the method TransportDeltaElectron(). Due to an electric field applied by a high voltage electrode at the middle of the TPC, these electrons then drift to a plane of anode wires, where the charges are amplified. Underneath the anode wires, the signal is read out on the pad plane. Electron-ion pairs, drift lines and electron avalanches can be visualized in Geant4. Figure~\ref{fig: Detector} shows track and drift lines created by a $1$~MeV electron inside the TPC.

\section{Geant4 simulation parameters and their optimization}
\label{Geant4 simulation parameters and their optimization}
For the extension of the Geant4 PAI model to include low energy primary particles~\cite{PAI_extension}, the deposited energy of these particles has been validated against experimental data. But when interfacing Geant4 and Garfield++, it is paramount that the simulation also produces the correct number of electron-ion pairs. In cases where the Geant4 PAI model is completely or partially responsible for the production (cases A,B,D), the correct setting of the lower production cut and the lowest electron energy limit is crucial.

\subsection{Lower production cut (LP-cut) and lowest electron energy limit (LEE-limit)}
\label{Lower production cut and lowest electron energy limit}
The lower production cut (LP-cut) is the minimum energy transfer required to produce a new particle. The creation of secondary particles is not possible if the transferred energy is lower than the LP-cut. In this case, the energy is transferred to the material, and no secondary particles are produced. The lowest electron energy limit (LEE-limit) on the other hand is the energy threshold below which an electron is not tracked anymore. If the kinetic energy of an electron falls below the LEE-limit value during a step, the full energy deposition is enforced,  independent of the material~\cite{lowestElectronEnergy}. All physics lists use a default value for the LEE-limit (\textit{e.g.} $100$~eV for the G4EMLivermore physics list), which can be modified by the user. Evidently, the two parameters influence only the production of secondaries and not the deposited energy, which remains stable.

\begin{figure}[htbp]
\centering
\subfloat[Mean number of electron-ion pairs\label{fig: mean}]{
\includegraphics[width=.495\textwidth]{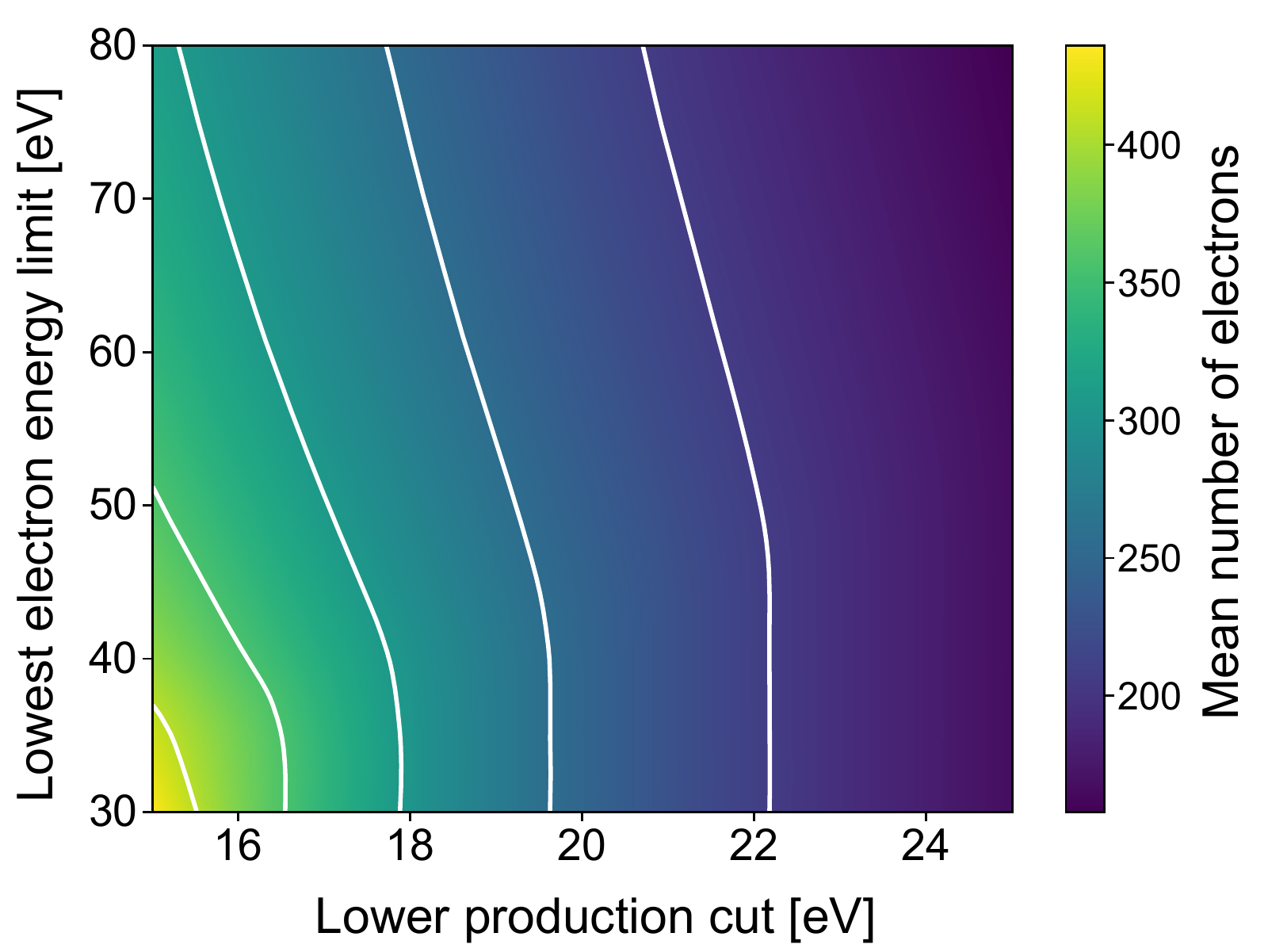}%
}
\subfloat[Variance of distribution\label{fig: variance}]{
\includegraphics[width=.495\textwidth]{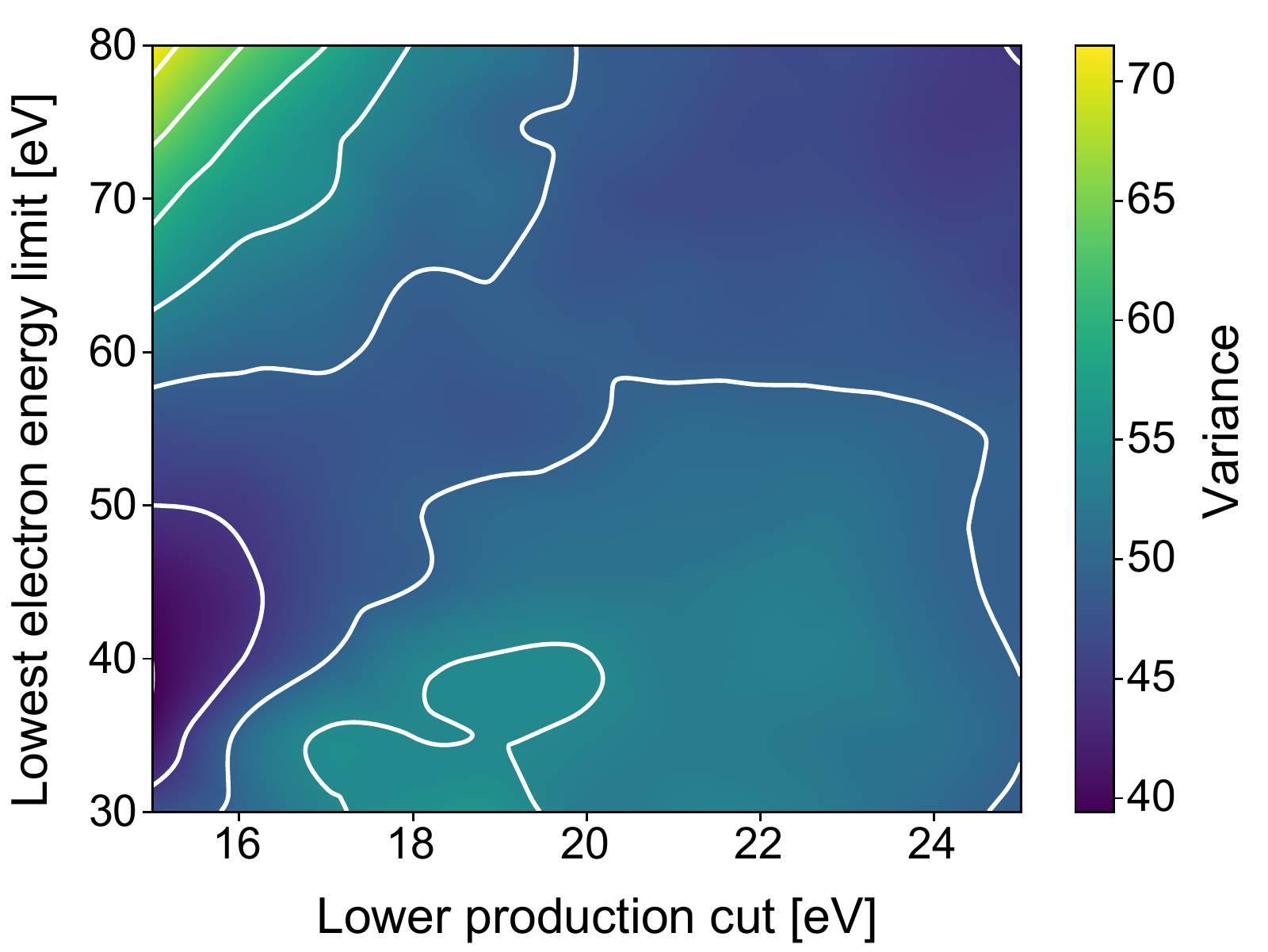}%
}
\caption{Geant4 PAI photon model simulation of the mean number of electron-ion pairs and the variance of the electron-ion pair distribution for a $10$~keV electron in a very large volume of He/isoButane $70$/$30$.}
\label{fig:Cut_LowELimit}
\end{figure}

Figure~\ref{fig:Cut_LowELimit} illustrates the influence of the LP-cut and the LEE-limit on the number of electron-ion pairs produced by the PAI photon model for a $10$~keV electron in a very large volume of He/isoButane $70$/$30$. As expected, figure~\ref{fig: mean} demonstrates that the mean number of electron-ion pairs rises with a decrease of the LP-cut and with a decrease of the LEE-limit. For LP-cuts larger than 15~eV, the mean number of electrons stays stable for a range of LEE-limits. The higher the LP-cut, the more stable is the number of electron-ion pairs. For the variance of the distribution on the other hand it is difficult to identify a clear behavior pattern for the two parameters, as indicates figure~\ref{fig: variance}. 

\subsection{W value and Fano factor}
\label{W value and Fano factor}
To determine the correct settings of the LP-cut and the LEE-limit, the deposited energy obtained in the simulations has to be compared to the number of electron-ion pairs multiplied by the W value. Additionally, it is desirable to obtain the correct Fano factor. The relevant definitions of W value and Fano factor are summarized by references~\cite{IAEA1995} and ~\cite{FanoFactors}, respectively. 

The W value, the mean energy needed to create an electron-ion pair, is defined as

\begin{equation}
\label{eq:W_value}
W(E) = \frac{E}{N},
\end{equation}

where $N$ stands for the mean number of electron-ion pairs produced by a particle of initial energy $E$, dissipating its complete energy in the gas volume. For high particle energies, \textit{i.e.} higher than $10$~keV for electrons, the W value reaches a constant asymptotic value $W_{a}$~\cite{IAEA1995}. Below $1$~keV electron energy, the cross-section ratio of ionizing processes versus non-ionizing processes becomes smaller compared to the high energy region. A certain fraction of the dissipated energy is carried by electrons that are too low in energy to further ionize. With $U$ representing the mean energy of these sub-ionization electrons, and $E$ the energy of the impinging particle, the energy-dependent expression for the W value can be written as 

\begin{equation}
\label{eq:WE_value}
W(E) = \frac{W_{a}}{1 -\frac{U}{E}}.
\end{equation}

For high-energy particles in thin absorbers, a differential w value 

\begin{equation}
\label{eq:w_value}
w = \frac{dE}{dN}
\end{equation}

is used. Since $W(E)$ reaches a constant value $W_{a}$ for energies of typically $10$~keV and higher, $w$ converges to the same constant value for sufficiently high energies. In case of full absorption, the kinetic energy of the primary particle is thus identical to the mean number of electron-ion pairs $N$ multiplied by the W value. For fast particles and thin absorbers, the number of electron-ion pairs N multiplied by the differential w value is equivalent to the deposited energy in the absorber. 

The production of pairs of charge carriers for a given energy loss is a correlated statistical process. The fluctuation of the electron-ion pair production with respect to the expectation from Poisson statistics is called the Fano Effect. For particles that are fully contained in an absorber, the Fano factor $F$ is defined as the ratio between the variance and the mean of the electron-ion pair distribution. 

\begin{figure}[htbp]
\centering
\subfloat[Work function W\label{fig: W}]{
\includegraphics[width=.495\textwidth]{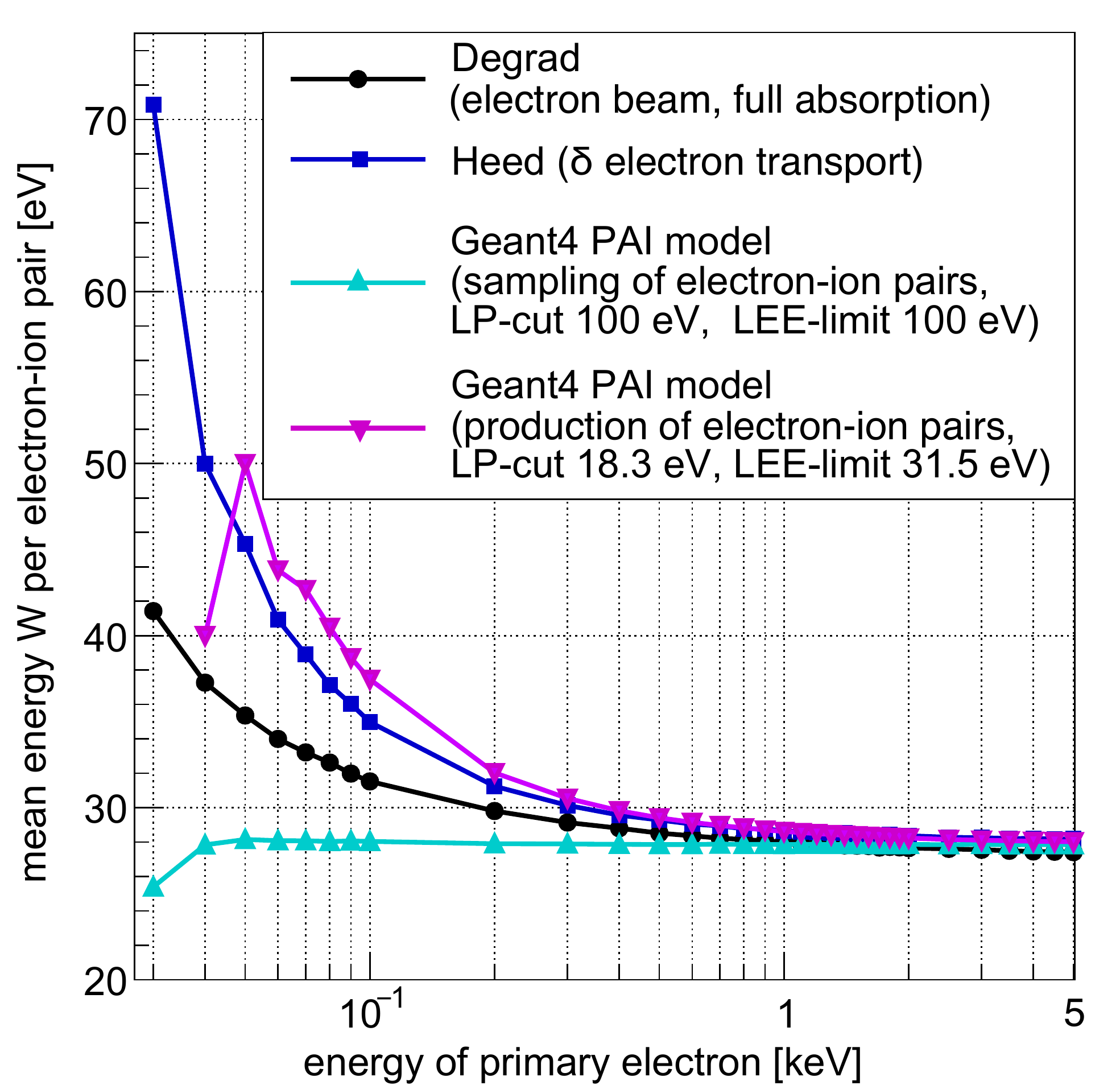}%
}
\subfloat[Fano factor\label{fig: Fano factor}]{
\includegraphics[width=.495\textwidth]{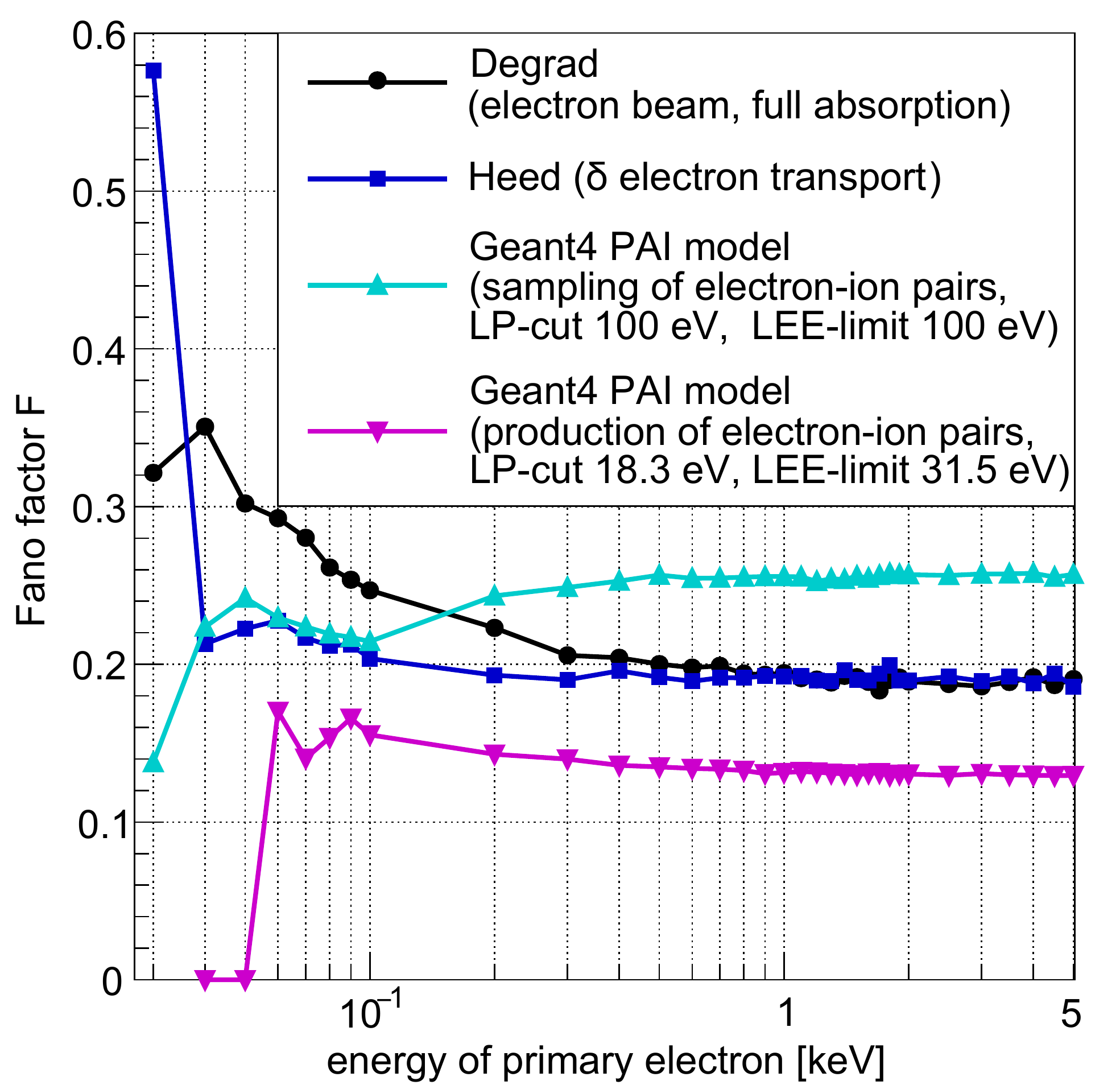}%
}
\caption{Simulated W value and Fano factor for $30$~eV to $5$~keV electrons in Ar/CO$_2$ $70$/$30$.}
\label{fig: W and Fano}
\end{figure}

Figure~\ref{fig: W and Fano} shows the W value and the Fano factor, simulated with four simulation options (Geant4 PAI model tuned to produce electron-ion pairs, Geant4 PAI model sampling electron-ion pairs, Heed and Degrad) for fully contained $30$~eV to $5$~keV electrons in Ar/CO$_2$ $70$/$30$.

To produce a sufficient number of electron-ion pairs, the settings of the Geant4 PAI model were modified to an LP-cut of $18.3$~eV and an LEE-limit of $31.5$~eV. The sampling of the electron-ion pairs depends only on the deposited energy per step, therefore the LP-cut and the LEE-limit where both left at $100$~eV. In Heed, the method TransportDeltaElectron() for $\delta$ electron transport was used to simulate the interaction of low energy electrons. For Degrad, the option \textit{electron beam, full absorption} was chosen for the simulation. In all four simulations, W reaches the correct asymptotic value of about $28$~eV for electron energies of $2$~keV and larger~\footnote{Whereas the W values for noble gases like argon (26 eV) and helium (41 eV) or quenchers like CO$_{2}$ (33 eV) and isobutane (23 eV) have been published(\cite{SauliW, Sharma}), the W values for gas mixtures are not easy to find. Most of the times they are calculated with programs like Heed from the W values of the components of the mixture using formula~\ref{eq:Wvalue_f}, which results in a W value of 28 eV for Ar/CO$_2$ $70$/$30$. More details about the W value calculation in Heed can be found in section~\ref{Optimal production cut for different gas admixtures}.}. For lower electron energies more energy is needed to create an electron-ion pair, therefore a higher W value is obtained. In spite of the parameter optimization, for electrons with a kinetic energy of less than $50$~eV, the Geant4 PAI model does not produce enough electron-ion pairs to determine the mean and variance of the distribution.

The small divergence between the results from Degrad, Heed PAI model and Geant4 PAI model (production mode), especially at low energies, can be explained by the approximations used in the calculations. The PAI model~\cite{Heed_PAI} relies on the assumption that a particle interacts only through the long-range dipole part of the electromagnetic force, and consequently uses an approximation similar to the Bethe-Bloch formula~\cite{BetheBLoch}. The dipole approximation of the force is valid to about 1$\%$ for energies down to typically $1$~keV for electrons or $2$~MeV for protons~\cite{private_communication2}. Below these energies, two problems arise. First, the Bethe Bloch formula requires phenomenological derived correction terms such as \textit{e.g.} Anderson-Ziegler\cite{andersen} and Lindhard-Scharff\cite{lindhard}. These terms are not included in the Heed PAI model. A second problem is that the non-dipole electromagnetic scattering becomes much larger below $1$~keV electron energy and approaches 60$\%$ of the dipole scattering at $80$~eV on atomic and molecular targets~\cite{private_communication1}. The Degrad program accounts for both dipole and non-dipole scattering and is, therefore, more accurate below $1$~keV electron energy. In case the electron-ion pairs are not produced by the Geant4 PAI model, but just sampled from the energy loss in a step, the W value is a non-energy dependent constant input parameter for the sampling function. The total simulated W value is thus also constant, and equivalent to the asymptotic W value. 

The Fano factor simulation shows the expected asymptotic behavior in Degrad and Heed. Starting with higher values for lower energies, a value of $0.19$ is reached for electron energies larger than $2$~keV. With the Geant4 PAI model (production mode) the correct Fano factor cannot be reproduced, the asymptotic value is too low at $0.13$. When sampling the electron-ion pair distribution from the energy deposit in a step, Geant4 uses a Gamma distribution as a generalization of a Poisson distribution with $\frac{N}{0.19}$ (mean number of electron-ion pairs per step divided by Fano factor) as parameter $k$ and $\frac{1}{0.19}$ as parameter $\lambda$. In spite of using the correct input parameters that should result in a Fano factor of $0.19$, the obtained Fano factor is too high with a value of $0.25$.
 
To summarize, figure~\ref{fig: W and Fano}~shows that Heed and Degrad reproduce the correct W value and Fano factor for fully contained electrons of $30$~eV to $5$~keV. With the correct settings of LP-cut and LEE-limit, the Geant4 PAI model (production mode) obtains the correct W value for electrons with a kinetic energy of at least $50$~eV. But it is not possible to simultaneously obtain the correct Fano factor since the variance of the electron-ion pair distribution seems to be too small. Based on these results, it seems to be the best option to use the $\delta$ electron transport of Heed for the tracking of low energy electrons with a kinetic energy of less than $5$~keV, and not the Geant4 PAI model.

\section{Verification of the Geant4/Heed PAI model interface}
\label{Verification}
Case D (Geant4/Heed PAI model interface) is the only case where different physics models are used for particle transport, \textit{i.e.} Geant4 and Heed, are used together. The PAI or PAI photon model of Geant4 is responsible for the primary ionization, and, following the recommendation of section~\ref{W value and Fano factor}, the TransportDeltaElectron() method of the Heed PAI model takes care of the $\delta$ electron transport to produce electron-ion pairs. Extensive simulations have been carried out to validate the results in comparison to the stand-alone simulation programs. All simulations in this paper have been carried out at Normal Temperature and Pressure (NTP) without the presence of an electromagnetic field.  G4EmLivermorePhysics was used as physics list with a fixed LEE-limit of 100~eV, and the PAI/PAI photon model was activated as additional EM physics model. For He/isoButane gas mixtures, the Geant4 10.4 PAI photon model was used, whereas the Geant4 10.3 PAI model was used for Ar/CO$_2$ gas mixtures. For Heed, the Garfield++ revision 541 was used. Since Heed is not multithread-processor safe, the Geant4 sequential mode has to be used for the time being.

\subsection{Lower production cut (LP-cut) and transfer energy threshold (TE-threshold)}
\label{Lower production cut and transfer threshold}
\begin{figure}[htbp]
\centering
\subfloat[W value\label{fig: W PAIphot}]{
\includegraphics[width=.495\textwidth]{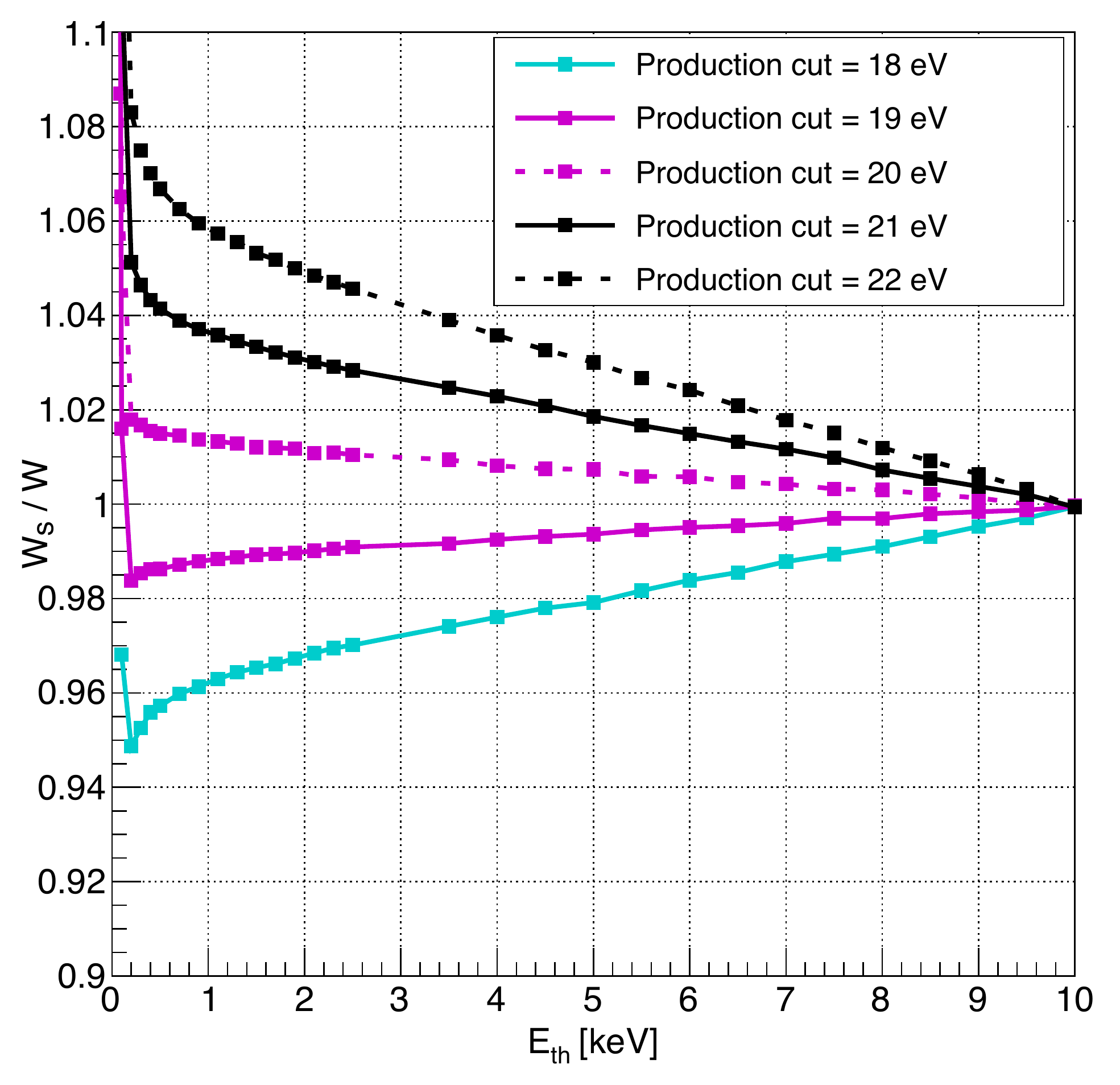}%
\label{fig:W_vs_Ethresh}
}
\subfloat[Fano factor\label{fig: F PAIphot}]{
\includegraphics[width=.495\textwidth]{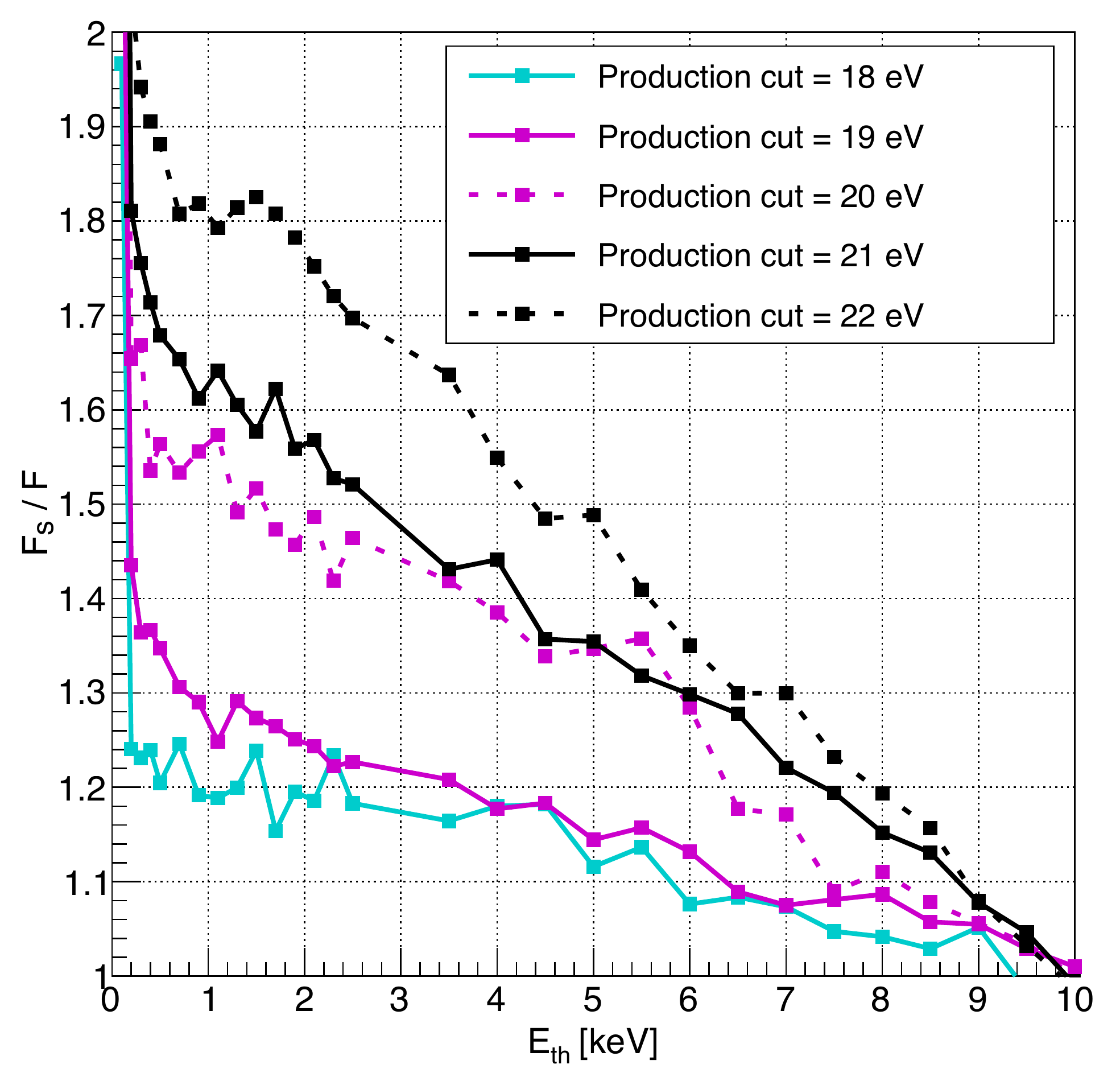}%
\label{fig:F_vs_Ethresh}
}
\caption{Influence of transfer threshold E$_{th}$ (TE-threshold) and lower production cut (LP-cut) on W value and Fano factor for 10~keV electrons using the PAI photon model in He/isoButane $70$/$30$. Figure (a) shows the comparison between simulated value W$_s$ and reference value W provided by Heed, and figure (b) the comparison between the simulated Fano factor F$_s$ and the Heed Fano factor.}
\label{fig: W and Fano depending on transfer threshold and lower production cut}
\end{figure}

When using the Geant4/Heed PAI model interface, the tracking of electrons is stopped in Geant4 when their kinetic energy falls below a certain energy threshold. The control is then handed over from Geant4 to Heed, where these electrons are re-created as $\delta$-electrons. The kinetic energy threshold, at which the electrons are transferred from Geant4 to Heed, is called transfer energy threshold (TE-threshold). The effect of using different LP-cuts and TE-thresholds, E$_{th}$, on the electron-ion pair distribution is shown in figure \ref{fig: W and Fano depending on transfer threshold and lower production cut}. For the simulations, the LEE-limit has been kept at the default value of $100$~eV for the G4EMLivermore physics model. 

For the W value (figure \ref{fig:W_vs_Ethresh}) and the Fano factor (figure~\ref{fig:F_vs_Ethresh}), the ratios between the simulated value and the Heed reference value are plotted~\footnote{The Heed reference values can be obtained directly from Heed. Using the EnableDebuggig() method of TrackHeed(), the W value and the Fano factor are printed to the screen. The Heed reference values do not depend on the transfer energy threshold E$_{th}$, since E$_{th}$ is only a parameter for the Geant4/Heed PAI model interface.}. For each of the plotted data points, $10^4$ electrons with an energy of $10$~keV were released in a sufficiently large gas volume, filled with a $70$/$30$ Helium-Isobutane mixture, to ensure full energy deposition. Subsequently, W was calculated using equation \ref{eq:W_value} and the corresponding Fano factor was obtained from the statistical mean and variance of the electron-ion pair distribution. Evidently, both the W value and Fano factor converge to the Heed values at a $10$~keV TE-threshold, as the electrons are immediately transferred to Heed. For TE-thresholds below $100$~eV, the simulated values rapidly increase towards infinity, as all electrons with a kinetic energy lower than this value are killed in the G4EMLivermore physics model. Looking at the W graph, an optimal LP-cut can be found somewhere between $19$ and $20$~eV, regardless of the TE-threshold. The Fano factor, on the other hand, follows a steady increase over the entire energy range. Whereas for the stand-alone Geant4 PAI model the Fano factor was too small, the values obtained with the interface seem rather too large. Whereas Heed has a database of literature values for W, a default value of $0.19$ for F is used for all gases. Hence, a value larger than $1$ in figure \ref{fig:F_vs_Ethresh} does not necessarily mean the interface produces a too large Fano factor. Additionally, a non-default Fano factor can be provided by the user in Heed, such that the desired values can be matched by the interface.

\begin{figure}[htbp]
\centering
\subfloat[W value\label{fig:W_10_1000}]{
\includegraphics[width=.495\textwidth]{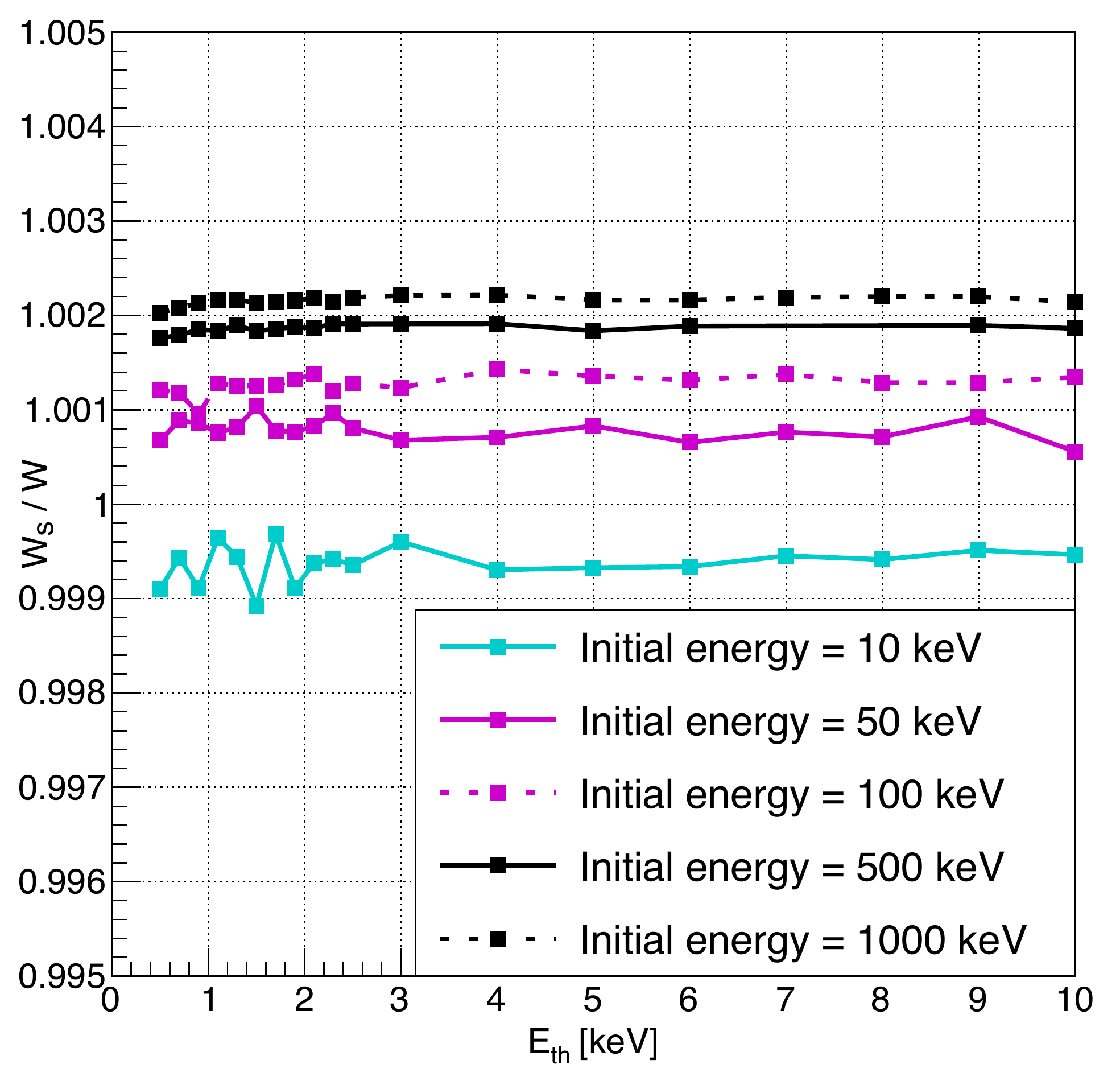}%
}
\subfloat[Fano factor\label{fig:F_10_1000}]{
\includegraphics[width=.495\textwidth]{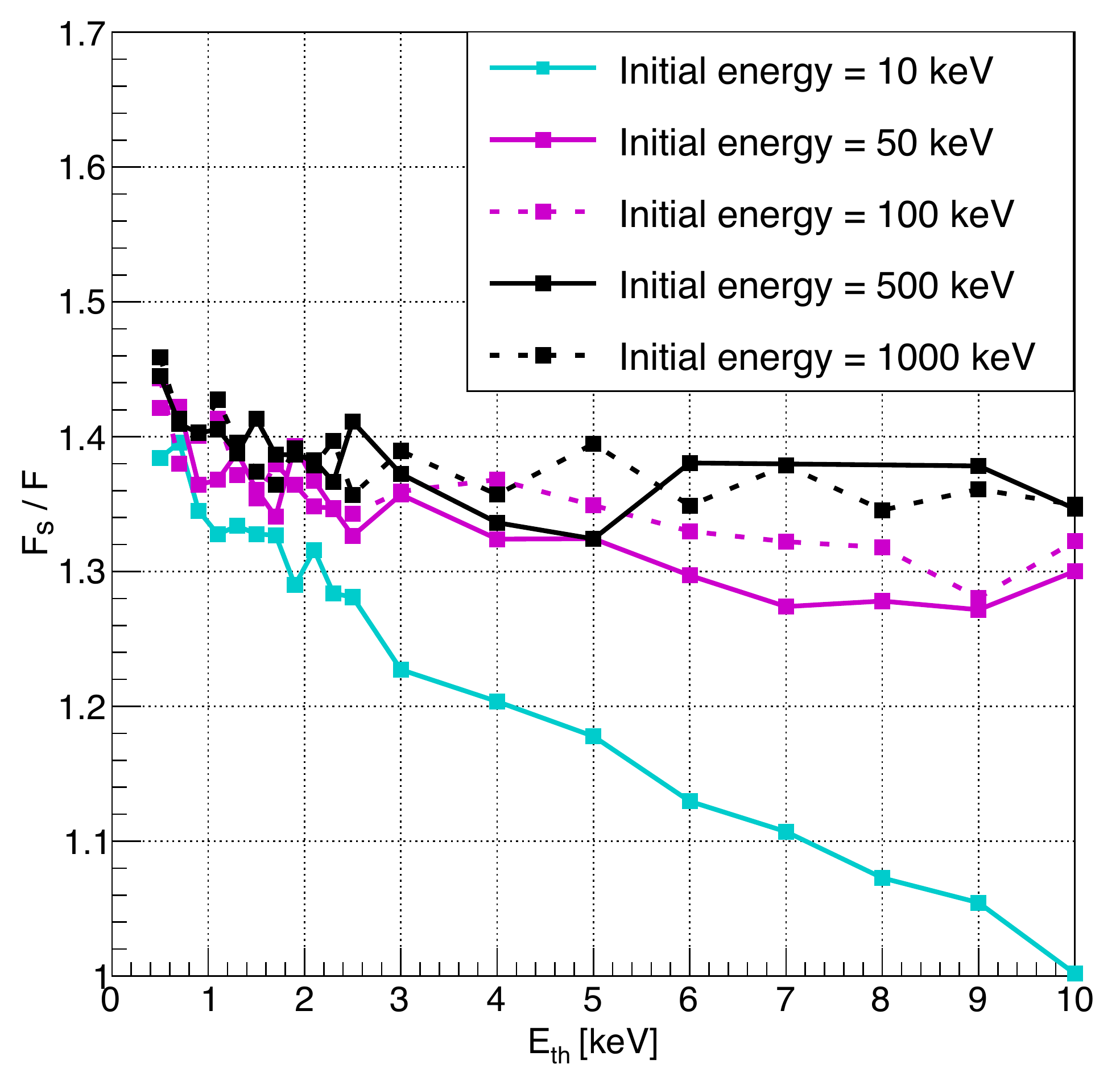}%
}
\caption{Comparative study of the W value and Fano factor for primary electrons between $10$~keV and $1000$~keV. An optimal production cut (LP-cut) of $19.45$~eV was used in all cases. Figure (a) shows the comparison between simulated value W$_s$ and reference value W provided by Heed, and figure (b) the comparison between the simulated Fano factor F$_s$ and the Heed Fano factor.}
\label{fig:WandFano_10_1000}
\end{figure}

Naturally, the results from figure~\ref{fig: W and Fano depending on transfer threshold and lower production cut} are futile if not the same optimal values are found for different energies of the primary. Using the optimal LP-cut of $19.45$~eV, in figure~\ref{fig:WandFano_10_1000} the W value and Fano factor were simulated for $10$~keV to $1000$~keV electrons. A relative difference of $\approx 0.1\%$ is observed for the obtained W values (figure~\ref{fig:W_10_1000}). The Fano factor, however, displays a clear discrepancy between the $10$~keV and the higher primary energies, except for TE-thresholds below $2$~keV, where the curves converge to comparable values. Therefore, it is recommended to set the TE-threshold to values between $100$~eV and $2$~keV when using the G4EMLivermore physics. For standard EM physics (LEE-limit of $1$~keV), it should be between $1$~keV and $2$~keV. 

\begin{figure}[htbp]
\centering
\subfloat[$1$~GeV particles\label{fig: particle types}]{
\includegraphics[width=.495\textwidth]{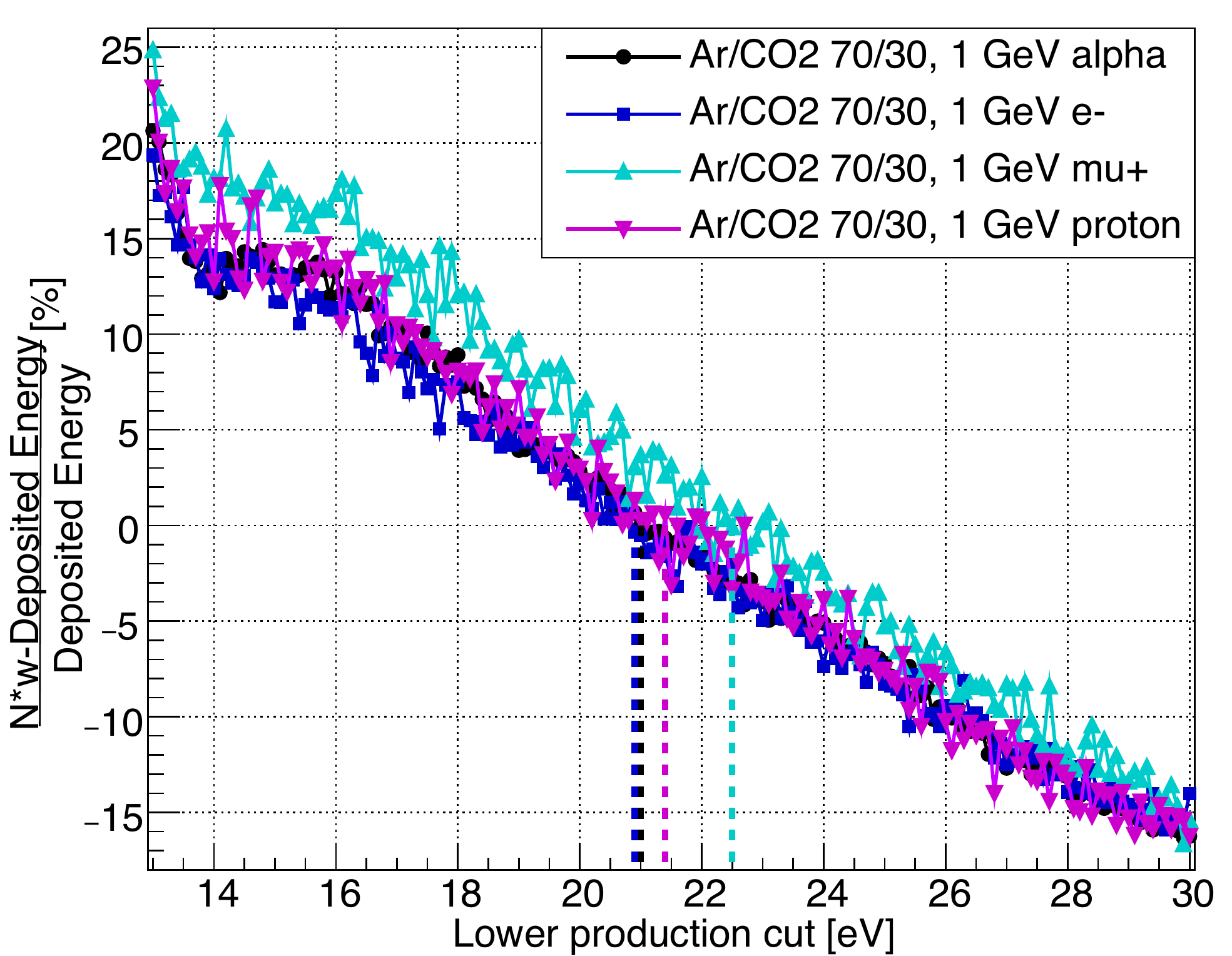}%
}
\subfloat[$10$~keV to $1$~GeV electrons\label{fig: energies}]{
\includegraphics[width=.495\textwidth]{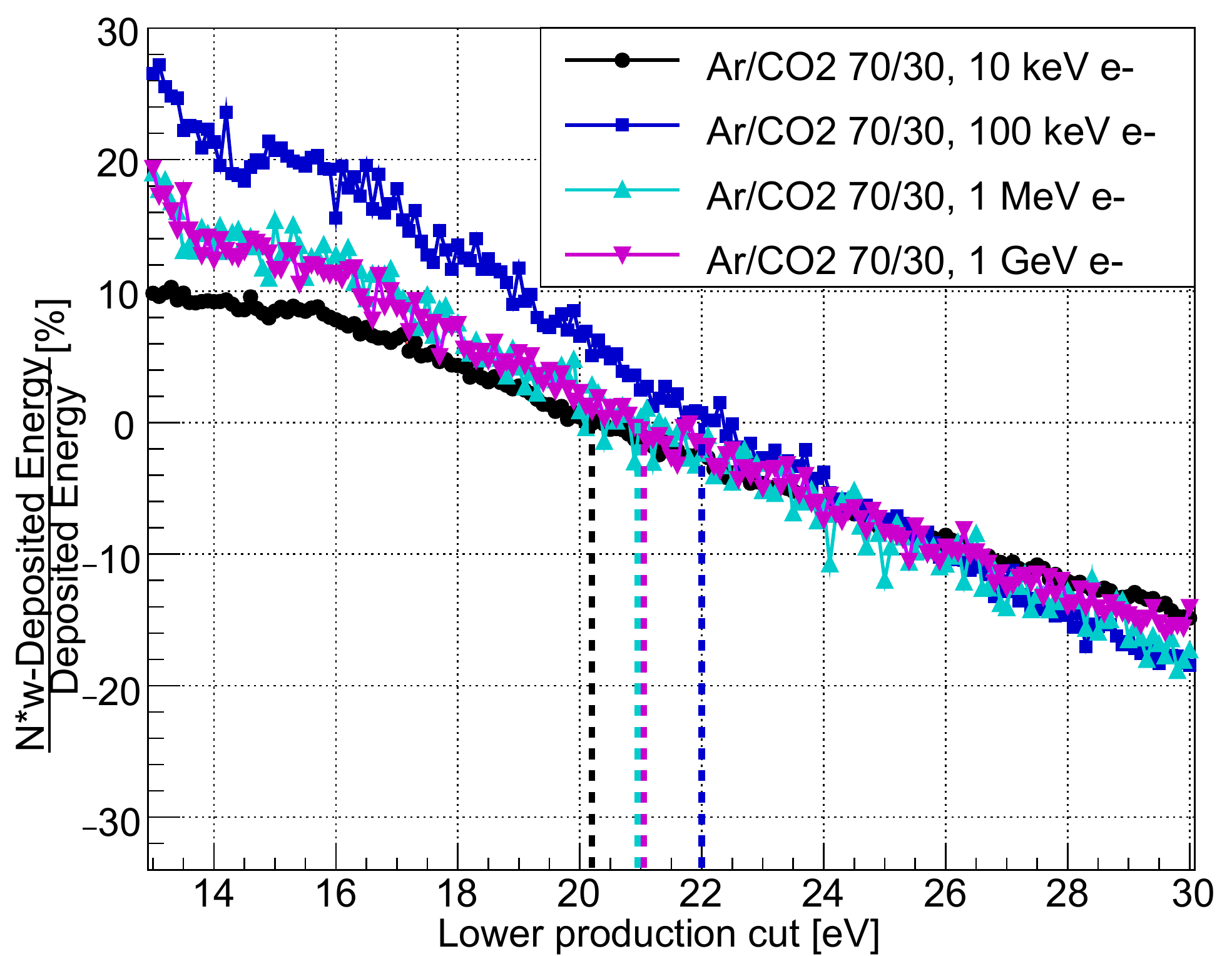}%
}
\caption{Different $1$~GeV particles and electrons of energies between $10$~keV and $1$~GeV in Ar/CO$_2$ $70$/$30$: Influence of the lower production cut (LP-cut) of the PAI model on the number of created electron-ion pairs in the Geant4/Heed PAI model interface. The figure shows the deviation of $N\cdot w$ from the deposited energy, simulated by Geant4, in percent.}
\label{fig: particle types and energies}
\end{figure}

The optimal LP-cut should also not vary much between different particle types and different particle energies for not fully contained particles. Figure~\ref{fig: particle types} displays the deviation of $N\cdot w$ (where n is the number of electron-ion pairs produced and w is the differential energy loss) from the simulated deposited energy in percent for $1$~GeV alpha particles, electrons, $\mu^+$ and protons in Ar/CO$_2$ $70$/$30$. Here, the energy dissipation of the primary particle is determined by Geant4 only, whereas the creation of secondary electrons is done by the interface, \textit{i.e.} both Geant4 and Heed. Again, the validity of the energy loss simulation by Geant4 is not considered here. The results of the interface, however, are verified by comparing $dE$ with $N\cdot w$ using equation \ref{eq:w_value}. For Ar/CO$_2$ $70$/$30$, at an LP-cut of $21$~eV, $N\cdot$w  of the Geant4/Heed PAI model interface is equivalent to the deposited energy of the PAI model in Geant4 for alpha particles and electrons. For $\mu^+$ and protons, the best production cut is slightly higher with $21.5$~eV and $22.5$~eV, respectively. Nevertheless, even at a cut of $21$~eV, N$\cdot$w for $1$~GeV proton is only $3 \%$ larger than the deposited energy. For the different electron energies of $10$~keV to $1$~GeV in figure~\ref{fig: energies}, the situation is similar. The best LP-cut for $1$~MeV and $1$~GeV electrons is $21$~eV, for $100$~keV electrons $22$~eV and for $10$~keV electrons $20$~eV. At $21$~eV, N$\cdot$w for $100$~keV electrons is only $3 \%$ higher than the deposited energy, whereas for $10$~keV electrons it is $2 \%$ lower. For simulations in Ar/CO$_2$ $70$/$30$, $21$~eV is therefore a good LP-cut. For other Argon and CO$_2$ gas admixtures, the best LP-cuts can be determined using the method shown in section~\ref{Optimal production cut for different gas admixtures}. 

To summarize, the optimization of the LP-cut and the TE-threshold for the Geant4/Heed PAI model interface is much easier than the tuning of the LP-cut in combination with the LEE-limit for option A (Geant4 production of electron-ion pairs). 

\subsection{Optimal lower production cut for different gas admixtures}
\label{Optimal production cut for different gas admixtures}
In a separate analysis, a parameterization of the optimal LP-cut as a function of mixture ratio was investigated. For this purpose, an approach similar to the determination of the W value of a gas mixture in Heed, was applied. Heed uses two assumptions to determine the W value. First, transfer reactions between atomic/molecular species, such as the Penning effect~\cite{FabioSauli2014}, are assumed to be negligible, which is not accurate for many mixtures with a small percentage of quenching gas~\cite{KrajcarBronic1996a}. Secondly, the total cross section for the interaction of a charged particle with a molecule or atom in the gas is assumed to be proportional to the total charge of that molecule or atom. Following this approach, the W value of a mixture is calculated as

\begin{equation}\label{eq:Wvalue_f}
W(f) = \frac{W_2 +(\frac{Z_1}{Z_2}W_1-W_2)f}{1+(\frac{Z_1}{Z_2}-1)f}.
\end{equation}

$W_1$ and $W_2$, and $Z_1$ and $Z_2$ are the W values and molecular charges of molecule $1$ and $2$, respectively, and f is the fraction of molecule $1$. As a first try, the same approach was used to determine the optimal production cut, resulting in the following expression

\begin{equation}\label{eq:prod_cut}
P(f) = \frac{P_2 +(\frac{Z_1}{Z_2}P_1-P_2)f}{1+(\frac{Z_1}{Z_2}-1)f},
\end{equation}

with $P_1$ and $P_2$ as the optimal LP-cuts for the two pure gases. This model was tested for different mixtures of Helium and Isobutane and mixtures of Argon and CO$_{2}$. The result of the simulations is shown in table \ref{tab:prod_cut}. The first column states the fraction for both He in He/isoButane and CO$_{2}$ in Ar/CO$_{2}$.

\begin{table}[H]
	\centering
	\caption{Optimal lower production cut (LP-cut) values for different gas fractions of the mixture.}
	\begin{tabular}{ccc}
		\toprule
		 & \multicolumn{2}{c}{production cut (eV)} \\
		\midrule
		$f$ & He/isoButane (He) & Ar/CO$_{2}$ (CO$_{2}$) \\
		\midrule
		$0$ & $18.71$ & $20.7$  \\
		$0.1$ & $18.74$ & $20.89$  \\
		$0.2$ & $18.79$ & $21.11$  \\
		$0.3$ & $18.84$ & $21.35$  \\
		$0.4$ & $18.92$ & $21.62$  \\
		$0.5$ & $19.01$ & $21.88$  \\
		$0.6$ & $19.17$ & $22.16$  \\
		$0.7$ & $19.45$ & $22.48$  \\
		$0.8$ & $20.07$ & $22.75$  \\
		$0.85$ & $20.86$ & -  \\
		$0.9$ & $21.95$ & $23.12$  \\
		$0.95$ & $25.19$ & -  \\
		$1$ & $30.54$ & $23.48$  \\
		\bottomrule
	\end{tabular}
	\label{tab:prod_cut}
\end{table}

\begin{figure}[htbp]
\centering
\subfloat[\label{fig:cut_heiso}]{
\includegraphics[width=.495\textwidth]{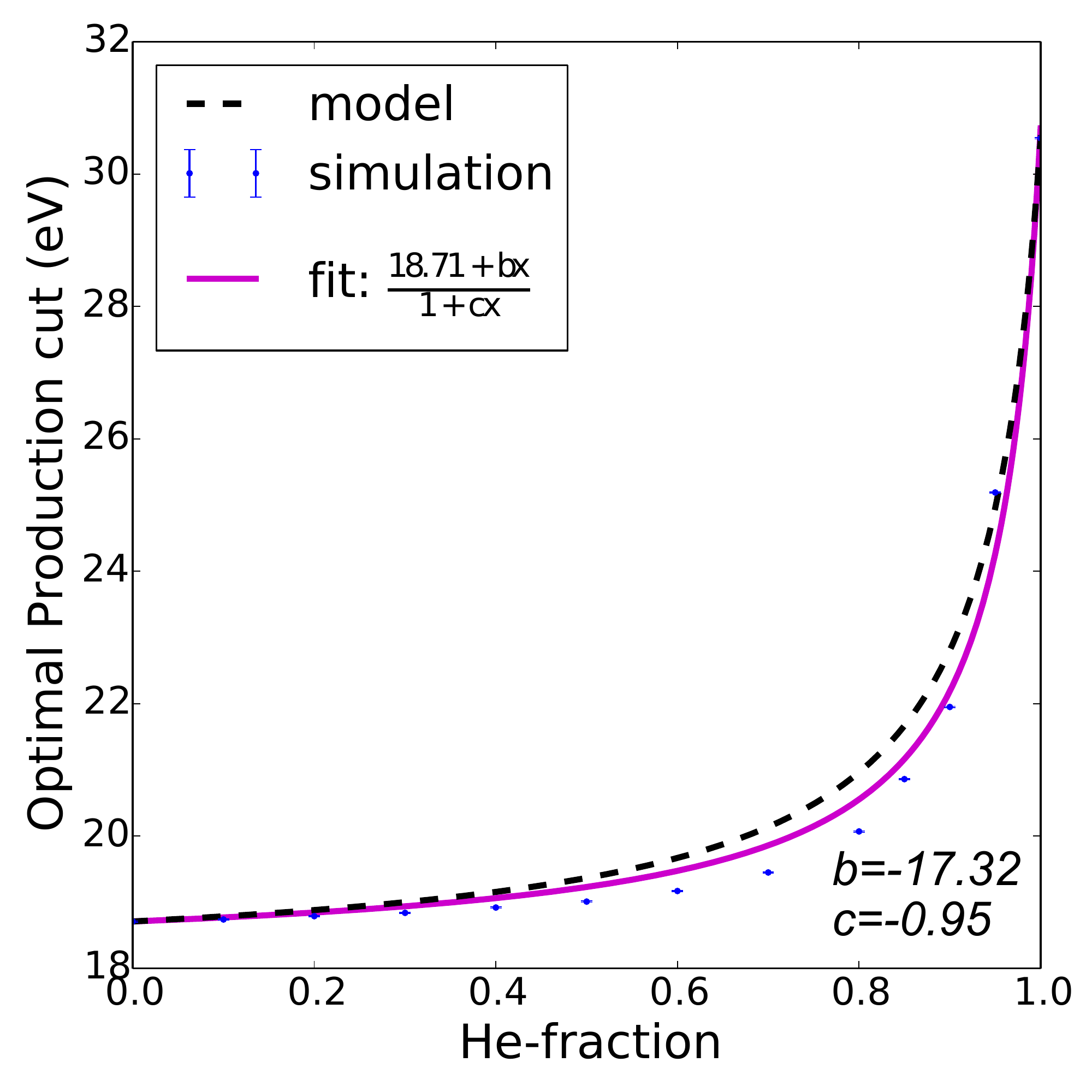}%
}
\subfloat[\label{fig:cut_arco2}]{
\includegraphics[width=.495\textwidth]{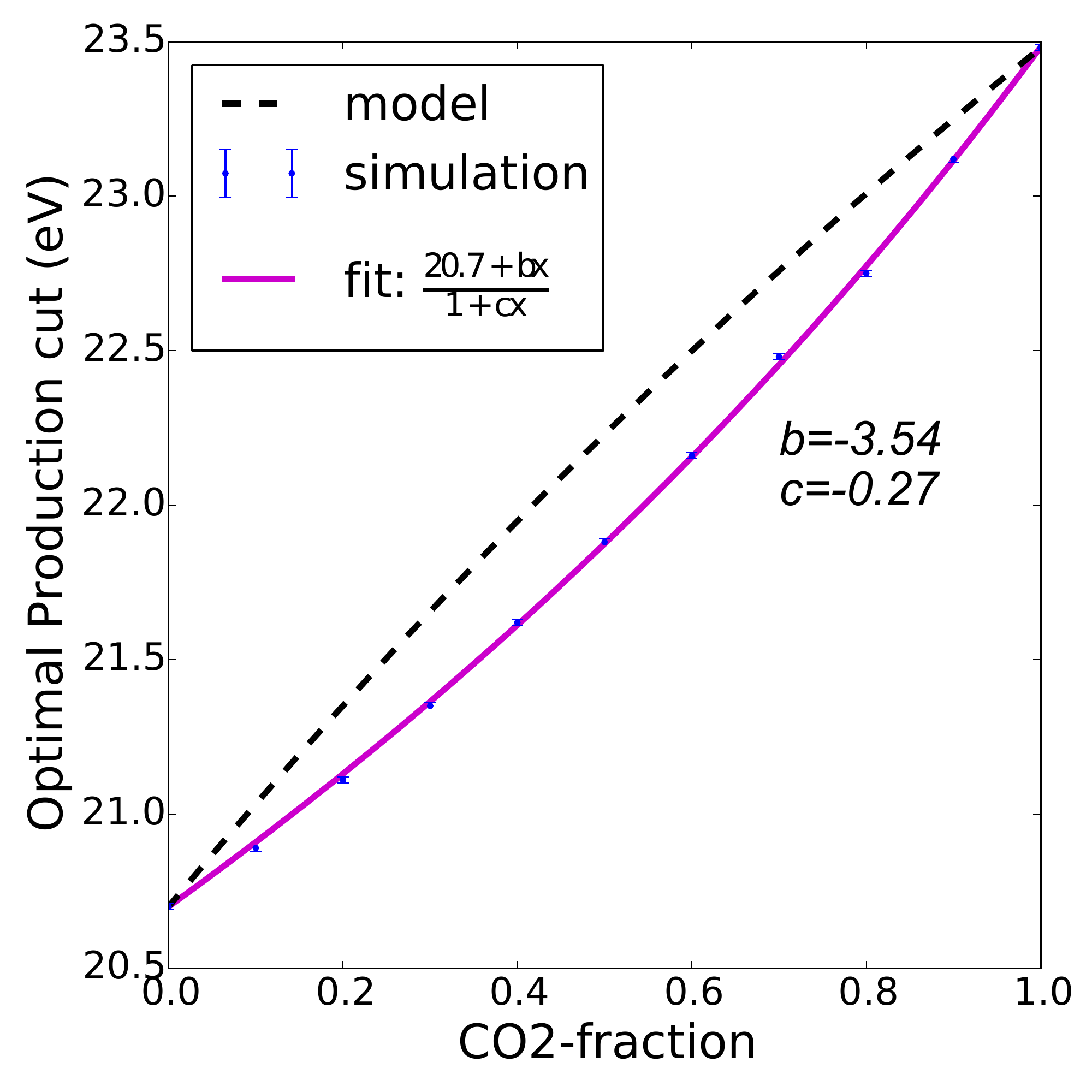}%
}
\caption{Comparison of the simulated data in table \ref{tab:prod_cut} and the model (equation~\ref{eq:prod_cut}) for the case of (a) He/isoButane and (b) Ar/CO$_{2}$. The dashed curve represents the model, the solid curve corresponds to the fit according to equation~\ref{eq:fit} and the error bars show the simulated data.}
\label{fig:opt_prod_cut}
\end{figure}

A comparison of the simulated data from table~\ref{tab:prod_cut} and the model described above is displayed in figure~\ref{fig:opt_prod_cut}. In the case of both He/isoButane (figure \ref{fig:cut_heiso}) and Ar/CO$_2$ (figure \ref{fig:cut_arco2}), the model clearly overestimates the data by a few percent. In a second attempt to parameterize the optimal LP-cut, the following function

\begin{equation}\label{eq:fit}
    P(f) = \frac{a+bf}{1+cf},
\end{equation}

was fitted to the data. In the fit, parameter $\mathbf{a}$ is fixed to the fraction $\mathbf{f=0}$ value from the table, and parameters $\mathbf{b}$ and $\mathbf{c}$ are left as free parameters. The resulting fit (red) shows a clear improvement over the model (green).  But the fitting approach requires additional effort, as a sufficient number of values need to be determined manually beforehand. Hence, the user may decide how precise W needs to be, and may even feed an experimentally determined W value to Heed, for which a production cut scan is necessary anyway.

\subsection{Deposited energy spectra}
\label{spectra}
\begin{figure}[htbp]
\centering
\subfloat[\label{fig: spectrum alpha}]{
\includegraphics[width=.495\textwidth]{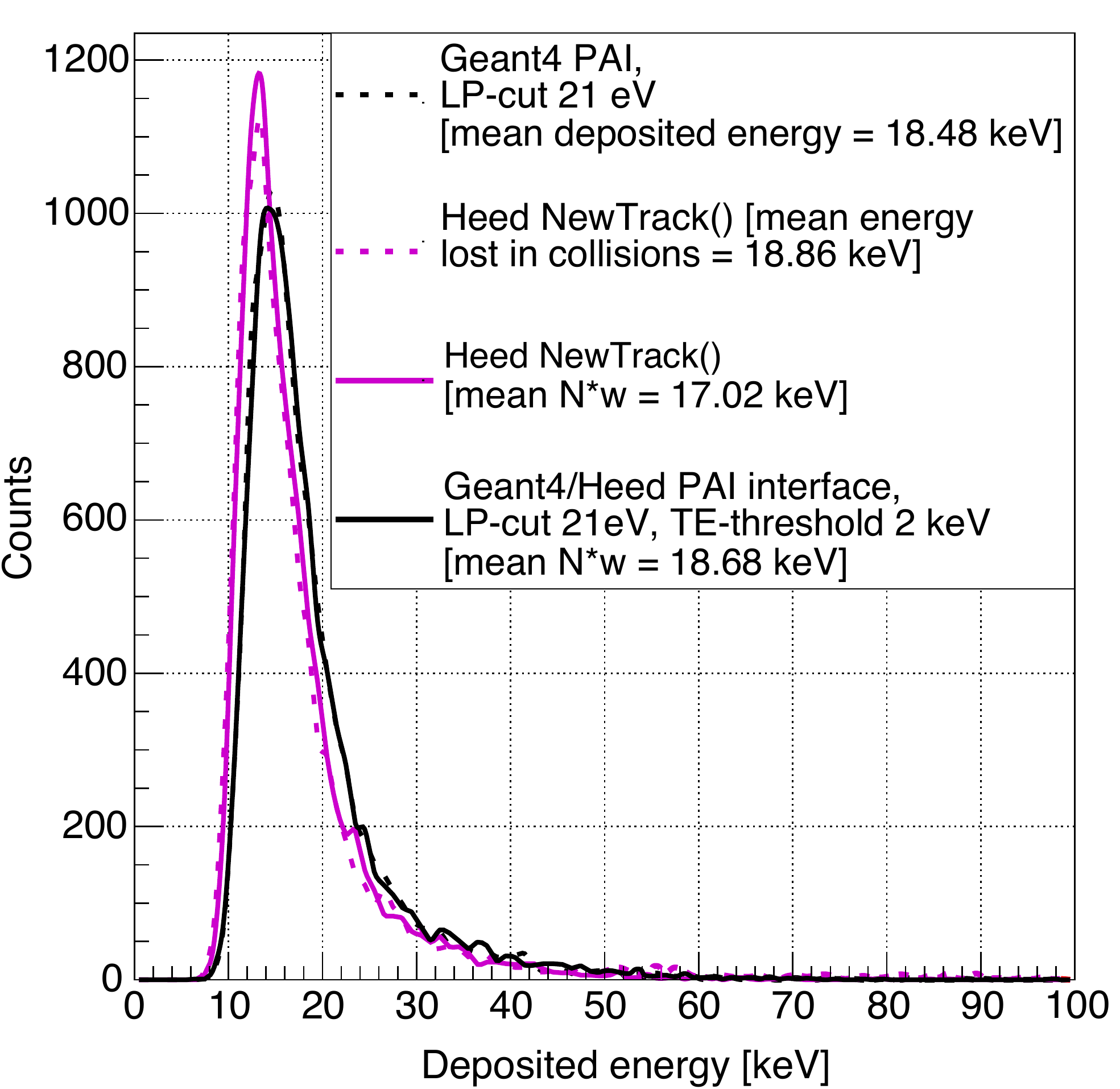}%
}
\subfloat[\label{fig: spectrum 1MeV}]{
\includegraphics[width=.495\textwidth]{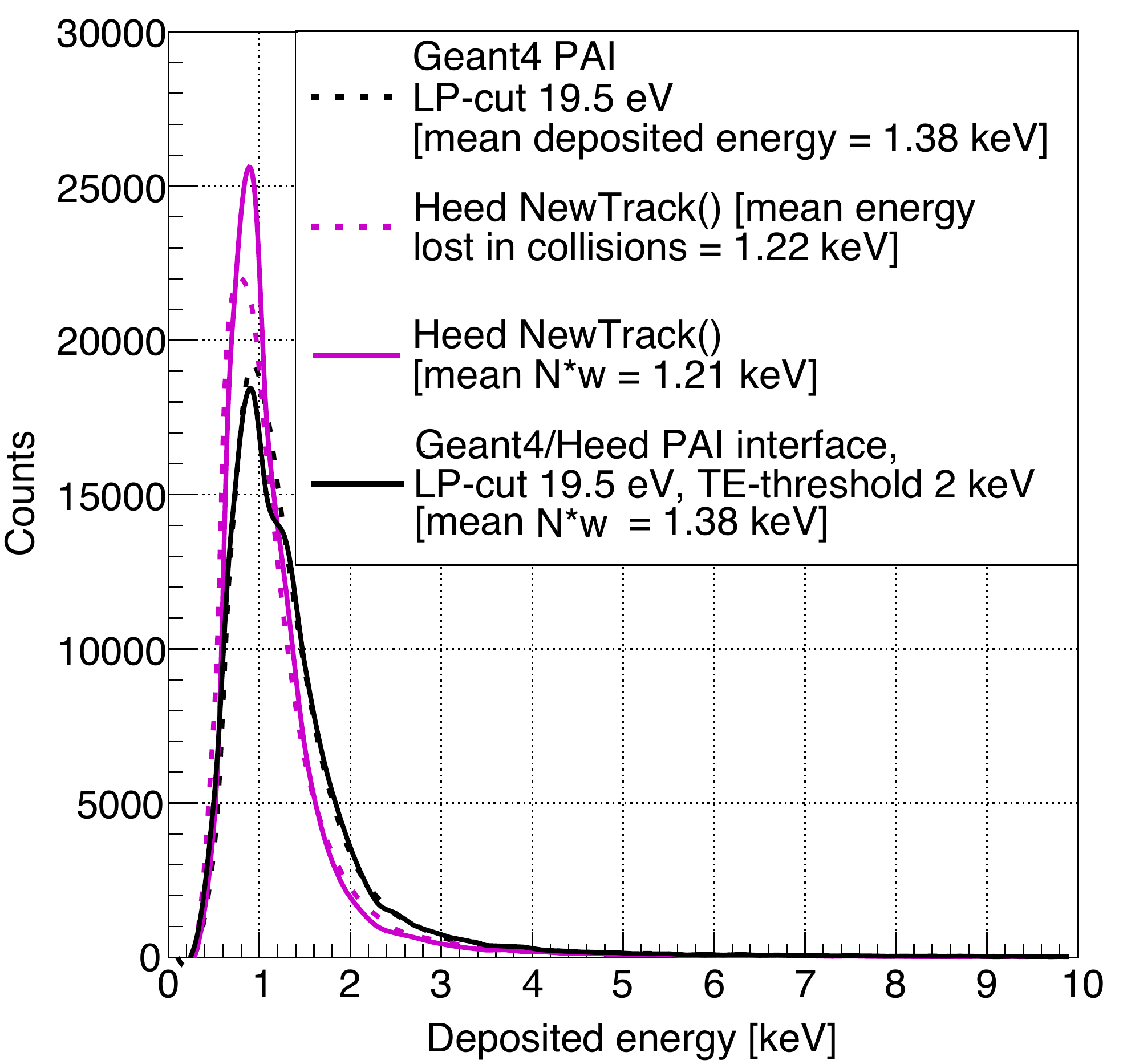}%
}
\caption{Relativistic particles: Deposited energy spectrum of (a) $1$~GeV alpha particles in Ar/CO$_2$ $70$/$30$ and (b) $1$~MeV electrons in He/isoButane $70$/$30$ in a $1$ cm thick layer. The black dotted curve represents the deposited energy simulated by Geant4. The purple dashed curve corresponds to the deposited energy calculated by Heed (NewTrack-method). The purple solid curve is showing the results for $N\cdot w$, $N$ being simulated by the NewTrack()-method in Heed and $w$ the literature value for $W\approx w$. The black solid curve displays the same quantity as the purple one, but simulated by the Geant4/Heed PAI model interface.}
\label{fig: spectra relativistic}
\end{figure}

In the case of particles not fully contained in the gas volume, it is interesting to study the deposited energy spectra. For all simulations, a TE-threshold of $2$~keV has been used. Figure~\ref{fig: spectra relativistic} displays the spectra for $10^{6}$ $1$~GeV alpha particles in $1$~cm of Ar/CO$_2$ $70$/$30$ and $10^{6}$ $1$~MeV electrons in $1$~cm of He/isoButane $70$/$30$. The black dashed line shows the deposited energy spectrum of the Geant4 PAI or PAI photon model, while the black solid line shows the deposited energy spectrum (N $\cdot$ w) simulated with the Geant4/Heed PAI model interface. Both curves have the same shape and the same mean value. From a Heed simulation using NewTrack(), the energy of the produced electrons clusters (purple dashed curve) and N $\cdot$ w (purple solid curve) have been obtained. In a volume of limited dimensions, here 10 cm x 10 cm x 1 cm, the cluster energy in Heed simulations is always larger than N $\cdot$ w, since some $\delta$ electrons in the clusters have a long range, and are not fully contained in the gas volume. With respect to the deposited energy spectrum, the Geant4 PAI model, the Geant4/Heed PAI model interface and Heed deliver comparable results for relativistic particles.

\begin{figure}[htbp]
 \centering
 \includegraphics[width=.6\textwidth]{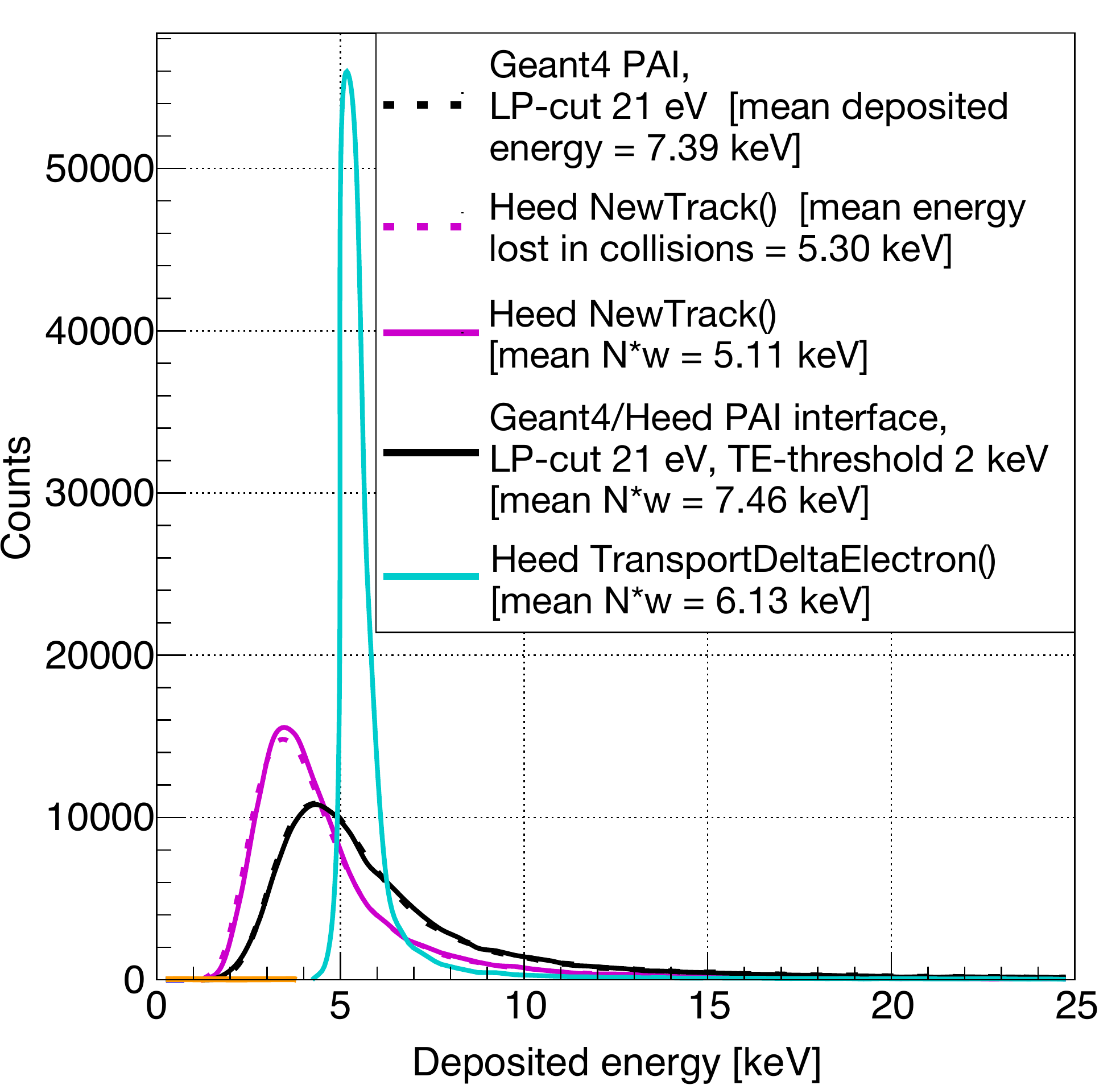}%
 \caption{Slow electrons: Deposited energy spectrum of $100$~keV electrons in Ar/CO$_2$ $70$/$30$.}
 \label{fig: spectra slow e-}
\end{figure}

For non-relativistic electrons, the situation is different. Figure~\ref{fig: spectra slow e-} shows the deposited energy spectra for $100$~keV electrons in  Ar/CO$_2$ $70$/$30$. Here, the Heed NewTrack() and the Heed TransportDeltaElectron() functions produce very different spectra, which both are incorrect and give a mean deposited energy that is too low~\footnote{The Heed TransportDeltaElectron() spectrum is missing the tail of higher energy depositions, and has thus a mean deposited energy that is too low, although the most likely energy deposition is too high.} The spectrum of the Geant4/Heed PAI model interface, on the other hand, is identical to the Geant4 PAI model spectrum.

\subsection{Spatial distribution of electron-ion pairs}
\label{Spatial distribution of electron-ion pairs}
The maximum energy of the $\delta$ electrons, which are sent to the interface, depends on the TE-threshold. A TE-threshold of $2$~keV is equivalent to selecting $\delta$ electrons with an energy of up to $2$~keV. Simulations of the range of a $2$~keV electron result in values of $160$~$\mu$m for Degrad, whereas Heed and Geant4 give $100$~$\mu$m and $70$~$\mu$m, respectively. Since Degrad includes the processes of photo-emission and absorption, it calculates a longer range than the other programs. To summarize, even with a transfer energy threshold of $2$~keV, the total range and the track shape of the simulated primary particle will be similar for the interface and a stand-alone Geant4 PAI model simulation due to the short range of the $\delta$ electrons. The track shape will be dominated by the distribution of the primary ionization electrons created by the PAI or PAI photon model in Geant4. 

\begin{figure}[htbp]
\centering
\subfloat[$1$~GeV e- in He/isoButane $70$/$30$\label{fig: position resolution 1GeV}]{
\includegraphics[width=.495\textwidth]{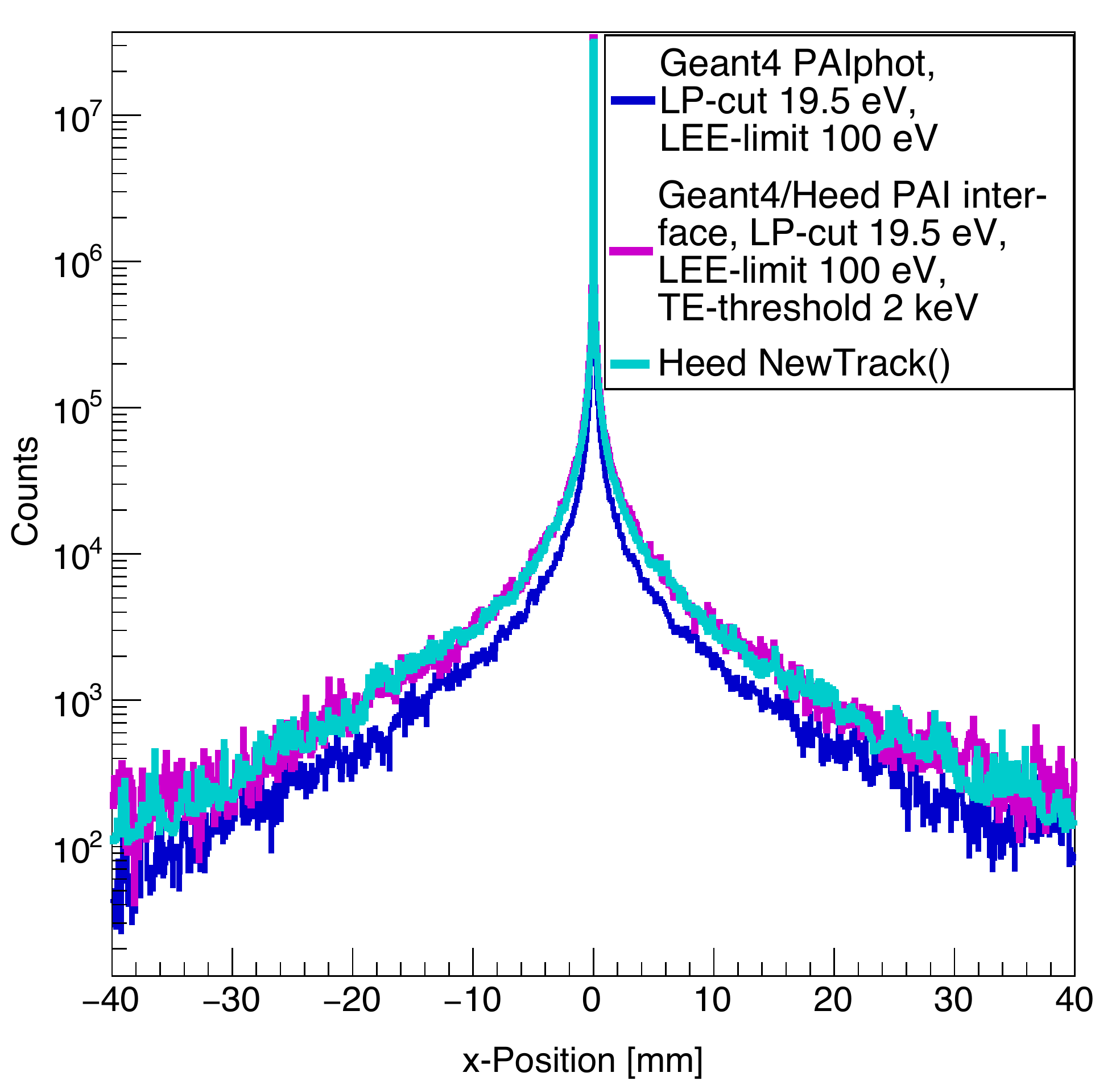}%
}
\subfloat[$100$~keV e- in Ar/CO$_2$ $70$/$30$\label{fig: position resolution 100keV}]{
\includegraphics[width=.495\textwidth]{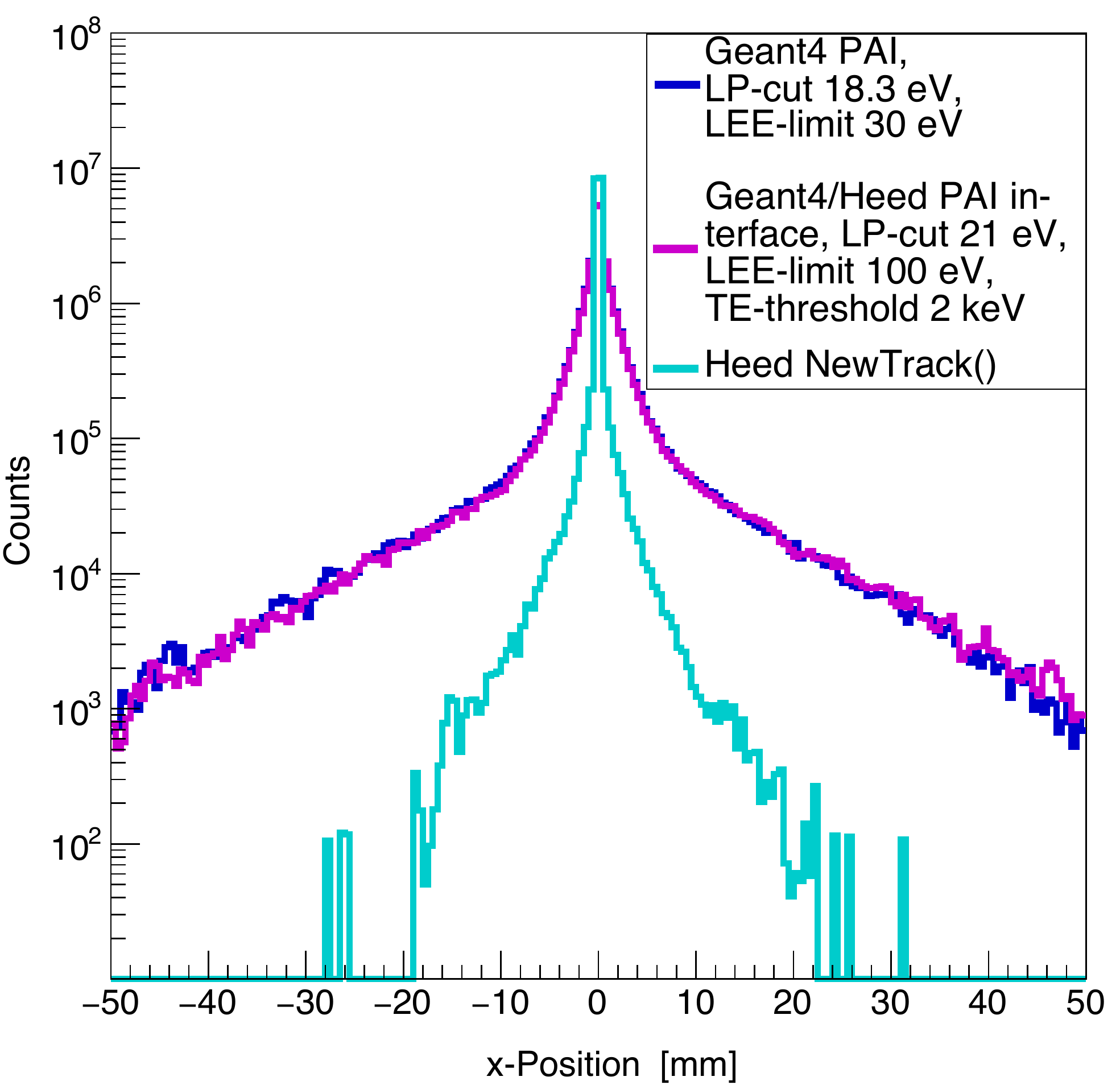}%
}
\caption{Spatial distribution of electron-ion pairs created by a $1$~GeV electron in He/isoButane $70$/$30$ and a $100$~keV electron in Ar/CO$_2$ $70$/$30$. The plot shows the x coordinate of the electron-ion pairs. The primary electron track starts in the origin of the coordinate system $(x_0,y_0,z_0)=(0,0,0)$, with the positive z-axis as momentum direction (axis pointing towards the reader).}
\label{fig: spatial distribution}
\end{figure}

Figure~\ref{fig: spatial distribution} shows the x position of all electron-ion pairs created by a $1$~GeV electron in He/isoButane $70$/$30$, and a $100$~keV electron in Ar/CO$_2$ $70$/$30$. The primary electron track starts in the origin of the coordinate system $(x_0,y_0,z_0)=(0,0,0)$, with the positive z-axis as momentum direction. For relativistic particles, the spatial distribution of the electron-ion pairs agrees nicely when comparing the Geant4/Heed PAI model interface and Heed as displayed in figure~\ref{fig: position resolution 1GeV}. For the Geant4 PAI photon model, the shape of the distribution is similar. Since the LEE-limit was kept at the standard 100~eV value of the G4EmLivermorePhysics, the Geant4 PAI photon model under-produces electrons. For $100$~keV electrons, on the other hand, figure~\ref{fig: position resolution 100keV} indicates that Heed creates a distribution of electron-ion pairs that is too narrow. This can be traced back to the absence of Coulomb scattering in the NewTrack()-method of Heed. The Geant4 PAI model and the Geant4/Heed PAI model interface (which uses the Geant4 PAI model instead of the Heed NewTrack() method) in contrast, do take this into account and therefore produce identical distributions. If one sets the LEE-limit to 30~eV and uses an LP-cut of 18.3~eV, as expected, the stand-alone Geant4 PAI model simulation does not under-produce electrons anymore, as visible in~figure~\ref{fig: position resolution 100keV}.

\subsection{Software Performance}
\label{Software Performance}
Regarding the software performance of the Geant4/Heed PAI model interface, the absolute time needed for a simulation depends of course on multiple factors, most importantly the performance of the computer hardware. Table~\ref{tab: performance} displays the performance of the following options: Geant4 PAI model, production of electron-ion pairs (case A), Geant4 PAI model, sampling of electron-ion pairs (case B), Heed PAI model (case C) and Geant4/Heed PAI model interface (case D). Since Heed is not multithread-processor safe, single-threaded programs were used. The program code was stripped down to include just the creation of the electron-ion pairs in the gas volume, not any drift or avalanche processes. The simulations were carried out on machine lxplus103 at CERN in direct successions, during a time where the authors were the only users of the machine. Equipped with a 10 2.4 GHz Intel Core Processors (Haswell, no TSX, IBRS), the machine has 30 GB of RAM and uses Centos7 as operating system. Averaged over 10 runs of 10$^{4}$ events, the Heed PAI model simulation (case C) is the fastest with 5.8 s. Very similar is the simulation time for the Geant4/Heed PAI model interface (case D) and the Geant4 PAI model, sampling of electron-ion pairs (case B) with 14.7 s and 17.9 s, respectively. Geant4 PAI model, production of electron-ion pairs (case A) is the slowest with 34.7 s, since the LEE-limit has to be set to $30$~eV, to obtain a sufficient number of electron-ion pairs. This slows the simulation down compared to the standard $100$~eV limit of the G4EmLivermorePhysics. In general, the Heed PAI model seems to be faster than the Geant4 PAI model, possibly due to Heed's missing Coulomb scattering in the NewTrack() method used for the primaries.

\begin{table}[H]
	\centering
	\caption{Comparison of simulation times for 10$^{4}$ 1~GeV electrons in 1 cm of Ar/CO$_2$ $70$/$30$.}
	\begin{tabular}{cc}
		\toprule
		Physics model in gas region & Time\\
		\midrule
		Geant4 PAI model, production of electron-ion pairs (case A) & 34.7 s \\
		Geant4 PAI model, sampling of electron-ion pairs (case B) & 17.9 s \\
		Heed PAI model (case C) & 5.8 s \\
		Geant4/Heed PAI model interface (case D) & 14.7 s \\
		\bottomrule
	\end{tabular}
	\label{tab: performance}
\end{table}

\section{Conclusion}
\label{Conclusion}
The present paper demonstrates how to interface Geant4 and Degrad or Geant4 and Garfield++ using the Geant4 parametrization feature. The aim is a complete simulation of gaseous detectors. There are several possibilities to divide the task between the software packages. Whereas the simulation of minimum ionizing particles (MIPs) is simple and can be carried out with Heed, the simulation of slow charged particles is more difficult. As discussed, the preferred way here is to use the Geant4/Heed PAI model interface. The PAI model in Geant4 creates ionization electrons, which are then transferred to Heed when their kinetic energy falls below $1$ to $2$~keV. To obtain the correct results with the Geant4/Heed PAI model interface, the tuning of the Geant4 lower production cut (LP-cut) is needed. Using the W value and the Fano factor, a method was developed to determine the optimal lower production cut for the PAI model for different gas mixtures. With the optimal LP-cut, the correct deposited energy spectra and correct spatial charge distributions are obtained for different particle types and energies. The software performance (simulation time) of the Geant4/Heed PAI model interface is better than that of stand-alone Geant4. For photons, the example of the xenon-based optical TPC demonstrates that complex physics cases can be simulated when interfacing Geant4 with Garfield++ and Degrad. To summarize, the technique of interfacing Geant4 with Garfield++ and Degrad with the help of the Geant4 parametrization feature is a very useful tools for detector simulations.

\section{Acknowledgments}
\label{sec:Acknowledgements}
This work was partially funded by the EU Horizon 2020 framework, BrightnESS project 676548.


\newpage
\bibliographystyle{elsarticle-num}
\bibliography{Geant4GarfieldInterface}

\end{document}